\documentclass[a4paper, 10pt]{article}
\usepackage{comment} 
\usepackage{geometry}
\newcounter{somecounter}
\usepackage{booktabs, array}
\usepackage{setspace}
\usepackage{float}
\usepackage{standalone}
\usepackage{amsmath,amsfonts,amssymb}
\usepackage{amsthm}
\usepackage{color}
\usepackage{xcolor}
\usepackage[english]{babel}

\usepackage{caption}
\usepackage{subcaption}
\usepackage{bm}
\usepackage{pdflscape}
\usepackage{authblk}
\usepackage{graphicx}
\usepackage{hyperref}
\usepackage[normalem]{ulem}
\usepackage{multirow}
\usepackage{multicol}
\usepackage[flushleft]{threeparttable}
\usepackage{enumitem}
\usepackage{url}
\usepackage[utf8]{inputenc}
\usepackage{dsfont}
\usepackage{kotex}
\usepackage{algorithm}
\usepackage{subcaption}
\usepackage{algpseudocode} % algorithmic
\usepackage{bbm}

\usepackage{lipsum}% http://ctan.org/pgk/lipsum

\usepackage{natbib}
\usepackage{mathtools}

\newcommand{\interior}[1]{%
 {\kern0pt#1}^{\mathrm{o}}%
}
\usepackage [english]{babel}
\usepackage [autostyle, english = american]{csquotes}
\MakeOuterQuote{"}

\usepackage{caption}
\usepackage[makeroom]{cancel}

\makeatletter
\newcommand*\bigcdot{\mathpalette\bigcdot@{.5}}
\newcommand*\bigcdot@[2]{\mathbin{\vcenter{\hbox{\scalebox{#2}{$\m@th#1\bullet$}}}}}
\makeatother

\title{\textbf{A Spatio-Temporal Dirichlet Process Mixture Model on Linear Networks for Crime Data}}

\author[1]{Sujeong Lee}
\author[3]{Won Chang}
\author[4]{Jorge Mateu}
\author[5]{Heejin Lee}
\author[1,2]{Jaewoo Park\thanks{Corresponding Author: Jaewoo Park, Department of Statistics and Data Science, Yonsei University; Department of Applied Statistics, Yonsei University, Seoul, 03722, Republic of Korea (e-mail: jwpark88@yonsei.ac.kr)}}

\affil[1]{Department of Statistics and Data Science, Yonsei University, Seoul, Republic of Korea.}
\affil[2]{Department of Applied Statistics, Yonsei University, Seoul, Republic of Korea.}
\affil[3]{Department of Statistics, Seoul National University, Seoul, Republic of Korea.}
\affil[4]{Department of Mathematics, University Jaume I, Castellón, Spain.}
\affil[5]{Department of Criminal Justice and Criminology, Sam Houston State University, Huntsville, USA.}

\begin{document}
\maketitle

\begin{abstract}
Analyzing crime events is crucial to understand crime dynamics and it is largely helpful for constructing prevention policies. Point processes specified on linear networks can provide a more accurate description of crime incidents by considering the geometry of the city. We propose a spatio-temporal Dirichlet process mixture model on a linear network to analyze crime events in Valencia, Spain. We propose a Bayesian hierarchical model with a Dirichlet process prior to automatically detect space-time clusters of the events and adopt a convolution kernel estimator to account for the network structure in the city. From the fitted model, we provide crime hotspot visualizations that can inform social interventions to prevent crime incidents. Furthermore, we study the relationships between the detected cluster centers and the city's amenities, which provides an intuitive explanation of criminal contagion. 
\end{abstract}

\noindent%

%%%%%%%%%%%%%%%%%%%%%%%%%%%%%%%%%%%%%
\section{Introduction}\label{sec:intro}

Research in criminology has shown that crime spreads as a contagious disease in local areas \citep{johnson2008repeat, mohler2011self}, with contagion rates varying between different types of crimes \citep{brantingham2021surge, brantingham2021gang}. The spread of crime dynamics in urban areas shows complex patterns due to the geographical structure of the city \citep{brantingham2009topology, dong2024spatio, song2013edge}. Therefore, considering the geographical structure can help us better understand crime dynamics, predict future incidents more accurately, and effectively develop prevention strategies. In this manuscript, we study the space-time locations of crime events in Valencia, Spain, recorded from 2018 to 2019, considering the city's road network structure. By using the exact space-time locations of crime events, we propose a novel spatio-temporal point process model that can understand the clustering patterns of the events.  

Point process models have been proposed to understand the clustering behavior of crime events. In this context, spatio-temporal Hawkes process models \citep{park2021investigating, reinhart2018review, zhuang2019semiparametric} can capture clustering and triggering patterns by modeling the intensity through the past history of the process. \cite{loeffler2018gun} developed a self-exciting spatio-temporal point process to model gun violence, analyzing diffusion and clustering by estimating spatial and temporal intensities. Bayesian inference was performed using an efficient Hamiltonian Monte Carlo method. \cite{lima2023hawkes} provides a comprehensive review of Hawkes point processes and their applications. Multivariate point processes have also been studied to model different crime types. \cite{hessellund2022semiparametric} developed a semiparametric regression model for the intensity function that depends on an unspecified factor common to all crime types. \cite{briz2023mechanistic} introduced a mechanistic bivariate spatio-temporal model to study the intensity of two crime types. \cite{mohler2014marked} proposed a marked point process model for multiple crime types, which can capture both short-term and long-term hotspots. However, these approaches have limited applicability to our particular data and motivating problem because they do not consider the road network structure. Our crime data can be naturally spatially inhomogeneous due to the network geometry, which additionally poses a significant challenge over classical spatio-temporal point process models. Specifically, if we do not consider the network structure, the places inaccessible to people could also be considered in the analysis. Furthermore, even if the two points are closely located in terms of Euclidean distances, the distance over the roads can be different depending on the network structure. Without considering such geometric structure, there can be non-negligible biases in the statistical inference.

There is a growing literature on computational methods for point processes on linear networks (see \cite{baddeley2021analysing} for a comprehensive review). Kernel density estimation methods have already been developed, such as the heat kernel \citep{mcswiggan2017kernel} that uses the density of a Brownian motion on the network. \cite{rakshit2019fast} proposed a convolution kernel estimator using a fast Fourier transformation. \cite{moradi2020first} developed summary statistics for point processes on a linear network, including the K-function and pair correlation function. \cite{d2023local} studied the local version of inhomogeneous second-order statistics for spatio-temporal point processes. \cite{rakshit2019efficient} developed a memory-efficient algorithm to compute second-order summary functions of point patterns on a linear network. Recently, in a more complex setting where marks are associated with each event, \cite{EckardtMoradi2024} developed summary statistics for various types of marks, including functional marks.

Although these previous works are useful for exploratory analysis, they do not employ model-based approaches. Developing point process models on road networks still remains challenging due to computational and inferential complexities. 
Recently, several spatio-temporal point process models have been studied in the context of linear networks. \cite{d2022inhomogeneous} proposed Gibbs point processes and log-Gaussian Cox processes to study the patterns of visitor stops by touristic attractions. \cite{d2022self} also developed spatio-temporal Hawkes point process models by including the network geometry in the inference procedure. They show that the proposed model fits much better compared to the planar one \citep{zhuang2019semiparametric}. \cite{gilardi2024nonseparable} developed a non-separable intensity function to model ambulance interventions on road networks. Apart from these contributions, no further model-based strategies have been proposed.

Although these approaches can describe spatio-temporal patterns of the process defined over linear networks, it is not trivial to detect the space-time clustering structure of crime events. Furthermore, fully Bayesian approaches have been under-explored. Such methods offer a natural framework for quantifying uncertainties within complex hierarchical models. A spatio-temporal point process typically has a hierarchical structure in which the observed process is modeled conditional on a latent process, and the latent process is modeled conditional on the model parameters \citep{cressie2011statistics}. Frequentist approaches typically estimate the model parameters first and then plug in these estimates to infer the latent process, ignoring parameter estimation uncertainty. This can underestimate standard errors, which becomes more severe as the latent structure becomes more complex. On the other hand, Bayesian methods construct a joint posterior of the parameters and the latent process. By drawing samples from this joint posterior, parameter and latent uncertainties are fully propagated.

In this line, we propose a spatio-temporal Dirichlet process (DP) mixture model to analyze crime events that occurred in the road network of the city. DP mixture models have been widely applied across various disciplines—for instance, to estimate the intensities of tree locations \citep{kottas2007bayesian}, to model crime patterns \citep{taddy2010autoregressive}, to cluster population genetic data \citep{reich2011spatial}, to model joint species distribution in ecology \citep{taylor2017joint}, and to detect disease hotspots \citep{park2023spatio}.

The main contributions of the manuscript are as follows. This study is the first attempt to develop a spatio-temporal DP point process on a linear network. Our model can automatically choose the number and location of cluster centers of crime events. To account for the structure of the linear network in the city, we adopt the convolution kernel estimator \citep{rakshit2019fast}, which can be quickly computed using the fast Fourier transformation. Second, we develop a fully Bayesian inference framework for the proposed model, enabling flexible hierarchical modeling and proper uncertainty quantification via posterior samples. Notably, most previous studies on point processes over linear networks have focused on exploratory analyses \citep{rakshit2019efficient, baddeley2021analysing, EckardtMoradi2024}, rather than formal statistical inference. Lastly, we introduce a model assessment approach that compares the empirical and theoretical proportions of events based on the likelihood function. We show that the proposed model more effectively captures spatio-temporal patterns in the crime data compared to the classical DP model.

With the proposed model, we analyzed two types of crime events—robbery and theft—recorded in Valencia from 2018 to 2019. Our main findings can be summarized as follows. First, crime clusters (i.e., hotspots) tend to influence within a spatial radius of approximately 1.1$km$ and a temporal window of about seven months. Furthermore, both robbery and theft exhibit peaks during the summer and fall seasons. Second, we observe that crime intensities are correlated with commercial areas, which tend to attract criminal activities. In addition, crimes are concentrated in urban, densely populated regions, whereas they are relatively rare in rural areas. Lastly, robbery and theft hotspots show an attractive tendency during peak periods. Moreover, theft occurrences have a positive effect on robbery and exhibit significant spatio-temporal autocorrelation.

The outline of the remainder of this paper is as follows. In Section~\ref{Data}, we describe the crime events data in Valencia, Spain. We also provide an exploratory data analysis for the dataset. In Section~\ref{Framework}, we propose a spatio-temporal DP mixture model on a linear network and describe details about model inference. In Section~\ref{Application}, we apply our method to real crime data, illustrating that our model can capture the spatio-temporal cluster centers of crime events. In addition, we study the relationship between cluster centers and the amenities of the city, which provides important sociological interpretations. In Section~\ref{Discussion}, we conclude the paper with a summary and some further discussion.

%%%%%%%%%%%%%%%%%%%%%%%%%%%%%%%%%%%%%
\section{Data description and exploratory analysis} \label{Data}

The dynamics of crime contagion are particularly complex within urban settings, influenced heavily by the geographic layout of the city. While crimes occur in a continuous space (\textit{e.g.}, within a city area measured by longitude and latitude), they are mostly confined to the street networks, influencing both the escape routes of criminals and the spatial distribution of crime. Early research also suggests that crime's contagious effects propagate along these street networks instead of dispersing freely, as evidenced by a fitted non-parametric influence kernel from real crime events. Motivated by these facts, in this section, we describe our crime data and provide some first exploratory data analysis.

\subsection{Data: crime events in Valencia, Spain} \label{data_description}

% \begin{figure}[htbp]
%     \centering
%     \begin{minipage}{0.25\linewidth}
%         \centering
%         \makebox{\includegraphics[width=\linewidth]{figure/agg obs entire.png}}
%         \subcaption{\label{fig:agg_obs}(a) Robbery}
%     \end{minipage}%
%     \hspace{0.05\linewidth}
%     \begin{minipage}{0.25\linewidth}
%         \centering
%         \makebox{\includegraphics[width=\linewidth]{figure/theft obs entire.png}}
%         \subcaption{\label{fig:theft_obs}(b) Theft}
%     \end{minipage}
%     \hspace{0.05\linewidth}
%     \begin{minipage}{0.28\linewidth}
%         \centering
%         \makebox{\includegraphics[width=\linewidth]{figure/amenity loc.jpg}}
%         \subcaption{\label{fig:amenity_loc}(c) Amenities in the city}
%     \end{minipage}
%     \caption{\label{fig:obs_and_amenity}
%     Locations of (a) robbery incidences, (b) theft incidences, and (c) amenities.}
% \end{figure}

\begin{figure}[htbp]
    \centering
        \centering
        \makebox{\includegraphics[width=\linewidth]{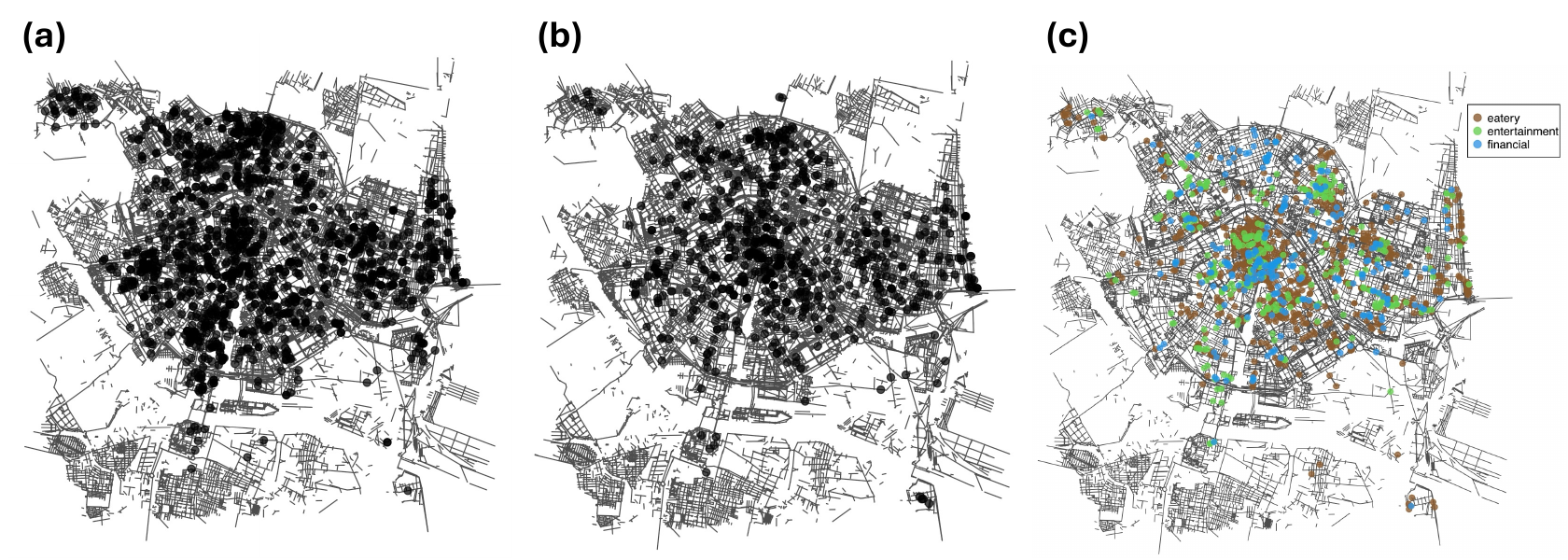}}
    \caption{\label{fig:obs_and_amenity}
    Locations of (a) robbery incidences, (b) theft incidences, and (c) amenities.}
\end{figure}

Valencia is a city on the Mediterranean coastline that has a population of more than 1.5 million \citep{dong2024spatio}. The crime data in this study is collected by the local police department in Valencia (Spain) and records the exact location and time for each crime event, allowing us to use the framework of spatio-temporal point processes. The raw dataset comprises four categories of crime: robbery, theft, alarms related to women, and other offenses. Robbery and theft are frequently examined in criminology due to their clear legal definitions and the substantial social harm they entail. For these reasons, we also focus on these two types of crime in our study. Robbery refers to the taking of property through force or threat, whereas theft involves the unlawful taking of property without consent. These definitions render both offenses more amenable to operationalization and measurement in crime statistics than more ambiguous or less detectable crimes (e.g., drug use, alcoholism, prostitution). Moreover, theft ranks among the most prevalent offenses globally \citep{vandijk2022}. Accordingly, robbery and theft are often employed in criminological research to test and refine theoretical frameworks, particularly those addressing property crime \citep{chastain2016, groff2007, johnson2008}. While robbery and theft may be linked to similar criminal opportunities (e.g., availability of suitable targets), resulting in overlapping geographical patterns, the two crimes are distinct—differing in the presence of violence, the level of risk to victims, and the situational dynamics under which they occur—and are therefore studied separately. Figure~\ref{fig:obs_and_amenity}a and \ref{fig:obs_and_amenity}b depicts the locations of these two crimes in the city. In our analysis, we used observations within 500 meters of each single street and projected them on the closest street, resulting in 1,423 records of robbery and 699 records of theft from 2018-01-01 to 2019-12-31 in the city of Valencia (Spain). We describe further details of the street network in Section~\ref{EDA}. We observe that the spatial patterns of both crimes are quite similar, occurring mainly towards downtown rather than in the outskirts of the city, with lots of variations in both the degree of the vertices and the lengths of the streets.

In addition to crime events, to investigate the relationship between the patterns of the reported crimes and the surrounding urban environment, we also have the geographical locations of different amenities in Valencia, including ATMs, banks, bars, cafes, nightclubs, pubs, and restaurants. 
In our study, we categorize the amenities into three groups: (a) entertainment (bars, nightclubs, pubs, with 286 amenities), (b) financial (ATMs, banks, with 246 amenities), and (c) eatery (cafes, restaurants, with 940 amenities). Figure~\ref{fig:obs_and_amenity}c illustrates the locations of these amenities in the city.

\subsection{Exploratory data analysis} \label{EDA}

Here, we set some background for point patterns on networks while providing some motivation for a cluster point process model. Consider a linear network \(\mathcal{L} = \bigcup_{i=1}^n l_i \subseteq \mathbb{R}^2\), as a finite union of line segments \(l_i \subseteq \mathbb{R}^2\) \citep{baddeley2021analysing}. The line segments are defined as \(l_i = [u_i, v_i] = \{ku_i + (1-k)v_i : 0 \leq k \leq 1\}\), where \(u_i, v_i \in \mathbb{R}^2\) are the endpoints of \( l_i \). For \( i \neq j \), the intersection of \( l_i \) and \( l_j \) is empty or an endpoint of both segments. When considering a time domain \(\mathcal{T} \subseteq \mathbb{R}^+\), we can introduce a spatio-temporal point process \(\mathbf{X}\) over the bounded domain \(\mathcal{L} \times \mathcal{T}\). This process thus contains information about the space-time locations for the crime events. 

We first convert Valencia's street network as a spatial object using \texttt{osmdata} \citep{osmdata} package in \texttt{R}. We can then represent the data through a projected coordinate reference system (EPSG code: 2062) of Spain that measures units in meters. Specifically, we select all the road features inside a rectangular window (with width 11.3$km$ and height 12.59$km$) defined by spatial coordinates, including the Valencia border. Then, we convert the data into a linear network format (i.e., \texttt{linnet} object in \texttt{spatstat.linnet} package). The linear network of Valencia inside a rectangular window contains 49,251 vertices (nodes) and 37,490 line segments (edges). However, using all line segments in the analysis is impractical because street crime events rarely occur on certain road types (e.g., highways), and analyzing the large network would involve high memory usage and computation cost. Therefore, we chose such streets where crime events could potentially occur, including residential streets, living streets, pedestrian streets, footways, service streets, steps, corridors, sidewalks, crossings, and cycleways. Residential streets serve as access to housing. Living streets are residential areas where pedestrians are legally prioritized, and vehicle speeds are strictly restricted. Footways are designated pathways primarily intended for pedestrian use. Service streets are roads within or leading to areas like alleys and business parks. Corridors are hallways within buildings. Following this procedure, we finally obtained a linear network consisting of 44,018 vertices (nodes) and 31,553 line segments (edges) with a width of 10.562$km$ and a height of 10.745$km$, referring to Valencia, whose total length of all line segments is about 1,992$km$.
%Specifically, we select all the road features inside a rectangular window (with width 10.95$km$ and height 13.73$km$) defined by spatial coordinates, including Valencia border. Then, we convert the data into a linear network format (i.e., \texttt{linnet} object in \texttt{spatstat.linnet} package). The linear network of Valencia contains 69,596 vertices (nodes) and 51,556 line segments (edges). However, using all line segments in the analysis is impractical because street crime events may not occur on specific road types (\textit{e.g.}, highways). Furthermore, it is computationally infeasible to consider the entire road network due to memory issues. Therefore, we chose such streets where crime events could potentially occur, including residential streets, living streets, pedestrian streets, footways, steps, corridors, sidewalks, crossings, and cycleways. Residential streets serve as access to housing.  Living streets are residential areas where pedestrians are legally prioritized, and vehicle speeds are strictly restricted. Footways are designated pathways primarily intended for pedestrian use. Corridors are hallways within buildings. Furthermore, we exclude the region where crime events are not observed. Following this procedure, we finally obtained a linear network consisting of 8,392 vertices (nodes) and 6,972 line segments (edges), referring to Valencia, whose total length of all line segments is approximately 805.94$km$. 

\begin{figure}[htbp]
    \centering
        \centering
        \makebox{\includegraphics[width=\linewidth]{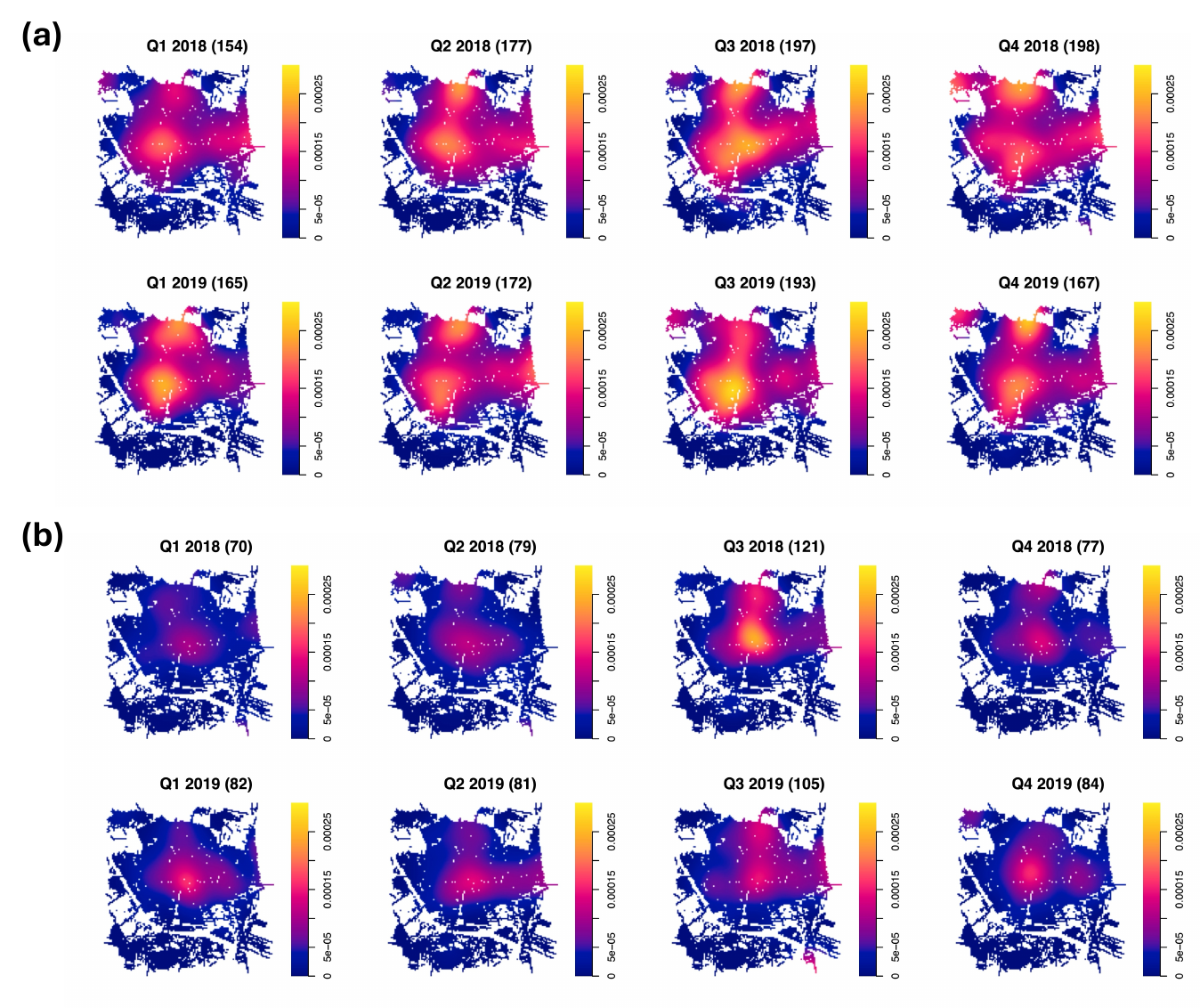}}
    \caption{Estimated first-order spatial intensity of (a) robbery and (b) theft by quarters and corresponding crime counts in the parentheses.}
    \label{fig:spatial_intensity}
\end{figure}

To understand the pattern of both types of crime events, we first compute first-order spatial intensities (depicted in Figure~\ref{fig:spatial_intensity}) by using the convolution kernel estimator \citep{rakshit2019fast} that accounts for the street structure. We use Scott's rule \citep{scott2015multivariate} to select the bandwidth for kernel density estimation. Figure~\ref{fig:spatial_intensity} illustrates that robbery crime events show higher intensity values in the top and bottom areas of the city, while theft crime events are more concentrated in the center of the city. These findings are aligned with Figure~\ref{fig:obs_and_amenity}. 

% \begin{figure}[htbp]
%     \centering
%     \makebox{\includegraphics[width=0.8\linewidth]{figure/figure3.pdf}}
%     \caption{\label{fig:kfunction}Estimated $K$-function for (a) robbery and (b) theft. Gray colors are simulation envelopes generated from homogenous Poisson processes, with the lower surface representing the minimum and the upper one representing the maximum $K$-functions. White colors indicate $K$-functions from the observed crime data.}
% \end{figure}
\begin{figure}[htbp]
    \centering
    \includegraphics[width=\linewidth]{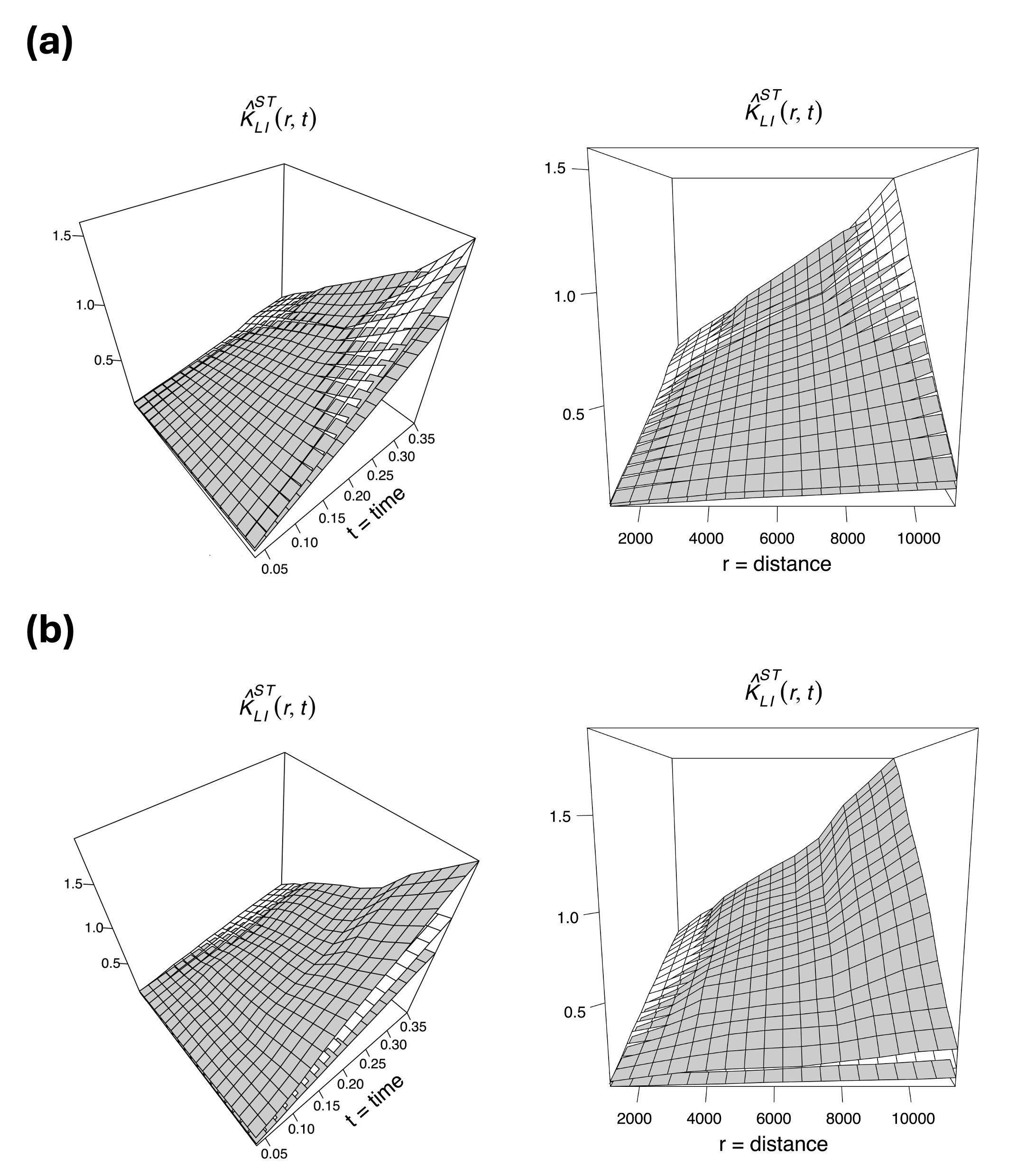}
    \caption{Estimated $K$-functions for (a) robbery and (b) theft, illustrated over time and distance dimensions. Gray colors are simulation envelopes generated from homogeneous Poisson processes, with the lower surface representing the minimum and the upper one representing the maximum K-functions. White colors indicate K-functions from the observed crime data. Areas where the observed $K$-function exceeds the simulation envelopes are highlighted in white.    
    }
    \label{fig:kfunction}
\end{figure}

We compute the linear network $K$-function \citep{moradi2020first} to study second-order characteristics of the point patterns. Specifically, the $K$-function is defined as 
% \begin{nalign}
%     K_{LI}^{ST}(r,t) = \mathbb{E} \left [ \sum_{(\mathbf{x},t_x)\in X} \frac{\mathbbm{1}\left\{ 0 < d_\mathcal{L}(\mathbf{u},\mathbf{x}) <r, |t_u-t_x| <t  \right\}}{\lambda(\mathbf{x},t_x) M((\mathbf{u},t_u),d_\mathcal{L}(\mathbf{u},\mathbf{x}), |t_u -t_x|)} |(\mathbf{u},t_u) \in X \right ], 
%     \label{eq:kfunction_edit}
% \end{nalign}
\begin{align}
    K_{LI}^{ST}(r,t) = \mathbb{E} \left [ \sum_{(\mathbf{x},t_x)\in X} \frac{\mathbbm{1}\left\{ 0 < d_\mathcal{L}(\mathbf{u},\mathbf{x}) <r, |t_u-t_x| <t  \right\}}{\lambda(\mathbf{x},t_x) M((\mathbf{u},t_u),d_\mathcal{L}(\mathbf{u},\mathbf{x}), |t_u -t_x|)} |(\mathbf{u},t_u) \in X \right ], 
    \label{eq:kfunction_edit}
\end{align}
where $\lambda(\mathbf{x},t_x)$ is the first-order intensity and $M((\mathbf{u},t_u),r,t)$ is the number of points lying exactly at the shortest-path distance $r$ and the time distance $t$ away from $(\mathbf{u}, t_u)$.
In (\ref{eq:kfunction_edit}), $d_\mathcal{L}(\mathbf{u},\mathbf{x})$ is shortest distance between $\mathbf{u}$ and $\mathbf{x}$ and $\left|\cdot\right|$ is the Euclidean distance.
% Following \cite{moradi2020first}, we obtain a nonparametric estimate of the $K$-function, $\hat{K}^{ST}_{LI}(r,t)$. Figure~\ref{fig:kfunction} shows that the observed $K$-functions exceed simulation envelopes, indicating that crime events (both robbery and theft) show a clustering behavior on certain space-time distances.
The $K$-function \citep{ripley1976second} has been widely used to assess whether a point pattern exhibits clustering, regularity, or complete spatial randomness. Specifically, point patterns are repeatedly simulated under a null model (i.e., complete spatial randomness), and the $K$-function is computed for each simulation. The K-function derived from the observed data is then compared to the range of simulated $K$-functions to detect significant deviations from the null hypothesis. This procedure is referred to as constructing simulation envelopes. In our study, we first compute a nonparametric estimate of the $K$-function \citep{moradi2020first}, $\hat{K}^{ST}_{LI}(r,t)$ from the observed point patterns. We then generate $K$-functions from 99 simulated homogeneous Poisson processes (i.e., the null model). When the observed $K$-function exceeds the simulation envelopes, it indicates significant clustering at the corresponding spatial and temporal scales. Figure~\ref{fig:kfunction} shows that crime events—both robbery and theft—exhibit clustering within a temporal range of 0.2 to 0.35 (approximately 5 to 8 months) and at spatial distances below 4,000 meters. In addition, robbery displays a notable clustering pattern at a temporal lag of around 0.15 (approximately 3 months) and at spatial distances exceeding 8,000 meters.
Following \cite{moradi2020first}, we also compute a p-value to test the null hypothesis of Poissonness (absence of any sort of clustering), defined as
% \begin{nalign} 
% T = \int\int \frac{\hat{K}^{ST}_{LI}(r,t)-E_K(r,t)}{\sqrt{V_K(r,t)}} drdt,\\
% \text{p-value} = \frac{1+ \sum_{i=1}^{m}\mathbbm{1}\{T_{i}>T^*\}}{m+1},
% \label{pval}
% \end{nalign}
\[T = \int\int \frac{\hat{K}^{ST}_{LI}(r,t)-E_K(r,t)}{\sqrt{V_K(r,t)}} drdt,\]
\begin{align}
    \text{p-value} = \frac{1+ \sum_{i=1}^{m}\mathbbm{1}\{T_{i}>T^*\}}{m+1},
\label{pval}
\end{align}
where $E_K$ and $V_K$ are the mean and variance of $\hat{K}^{ST}_{LI}(r,t)$. In \eqref{pval}, each $T_i$ is computed from a simulated point pattern generated from a homogeneous Poisson process (i.e., null hypothesis), and $T^*$ is computed from the observed point pattern. The p-values for robbery and theft were 0.02 and 0.04, respectively; therefore, we can conclude that the crime events are not compatible with a homogeneous spatio-temporal Poisson process and, indeed, highlight structures with varying clustering degrees.

%%%%%%%%%%%%%%%%%%%%%%%%%%%%%%%%%%%%%
\section{A Dirichlet process mixture model on a linear network} \label{Framework}

We propose and develop a spatio-temporal DP mixture model considering the geometrical structure of the linear network. The method can detect unobserved cluster centers of street crimes and underpin hotspots. 

\subsection{Model framework}
\label{modelframework}

Let $\lbrace (\mathbf{x}_i,t_i) \rbrace_{i=1}^{N}$ be a realization of the point process $\mathbf{X}$ over the bounded domain $\mathcal{L}\times\mathcal{T}$. We consider the space domain over the linear network $\mathcal{L} \in \mathbb{R}^2$, and a normalized time domain $\mathcal{T} = [0, 1]$. Normalizing the time domain has several practical advantages and is commonly adopted in spatio-temporal point process analysis \citep{moller2003statistical,baddeley2016spatial}. First, the normalization removes the influence of specific time units (e.g., weeks, months, years) on model fitting, allowing for consistent interpretation of model results across different temporal scales. Furthermore, rescaling the temporal domain can improve the computational stability of MCMC algorithms. In our problem, each $(\mathbf{x}_i,t_i)$ denotes the location and time of a crime event. We introduce the latent space-time cluster centers as $\mathbf{c}=\{(\mathbf{c}_j^s, c_j^t)\}_{j=1}^M \in \mathcal{L} \times \mathcal{T}$. To assign an $i$th observation to a cluster, we define the cluster membership variable $g_i \in \{1, \ldots, M\}$ and model the membership variable as a DP prior \citep{ishwaran2001gibbs}. Specifically, we model a $g_i$ as
\[
g_i \sim \text{Categorical}(q_1, \ldots, q_M),
\]
where \(q_1, \ldots, q_M\) represents the cluster probabilities for crime events. Then we model the cluster membership probabilities as the stick-breaking process prior \citep{sethuraman1994constructive}:

\begin{align}
q_{j} = \begin{cases} 
U_1, & \text{for } j = 1 \\
U_j \prod_{m=1}^{j-1} (1 - U_m), & \text{for } j = 2, \ldots, M,
\end{cases}
\label{eq:stick-breaking}
\end{align}
where \(U_1, \ldots, U_m \overset{\text{iid}}{\sim} \text{Beta}(1, b_u)\) with rate parameter \( b_u \). In our analysis, we set a hyperprior for $b_u$ as $\text{Gamma}(1, 1)$ and $\text{Gamma}(1, 1/4)$, following \cite{reich2011spatial}. Note that $b_u$ controls the expected number of clusters. This choice can encourage the number of non-empty clusters to lie between 1 and 15, thereby allowing the regularization to reduce the possibility of overfitting. In the stick-breaking process prior, the initial mixture probability $q_1$ is represented as a beta random variable $U_1$. For $j = 2, \cdots, M$, the subsequent mixture probabilities $q_j$ are obtained by multiplying $1 - \sum_{m=1}^{j-1} q_m $, the remaining probability, with $U_j$, the proportion assigned to the $j$th cluster component. Theoretically, \eqref{eq:stick-breaking} considers an infinite number of clusters. However, in practice, we approximate it with a sufficiently large $M < \infty$. The remaining probabilities $1 - \sum_{m=1}^{j-1} q_m$ become 0 for large $j$; therefore, no observation is assigned to the $j$th cluster. From this finite representation, we can automatically choose the number of non-empty cluster centers. 

After we assign the cluster membership to each observation, we model the space-time locations of crime events. Specifically, we define the conditional distribution for a given $g_i$ as
\begin{align}
f((\mathbf{x}_i, t_i) \mid g_{i} = j)
& \propto K_S(\mathbf{x}_i;\mathbf{c}_j^s, w_{s})K_T(t_i; c_{j}^{t}, w_{t}),
\label{eq:spatialkernel}
\end{align}
where $K_S(\cdot)$ and $K_T(\cdot)$ are spatial and temporal kernels, respectively. In our study, we use $K_T(t_i; c_{j}^{t}, w_{t})$ as a Gaussian kernel with center $c_{j}^{t}$ and a standard deviation $w_{t}$ (temporal range of cluster centers). There are several options for the spatial kernel $K_S(\cdot)$ that can account for the network geometry (see \cite{baddeley2021analysing} for a comprehensive review). One might consider the equal-split continuous kernel estimator \citep{okabe2012spatial}, which is a path enumeration method. At each fork in the linear network, the method divides the remaining tail mass at the vertex equally across the outgoing line segments. Therefore, the method can preserve the total mass. However, the implementation of the equal-split continuous kernel estimate is computationally demanding for large observations with complex street structures, as happens in our case. The heat kernel \citep{mcswiggan2017kernel}, which exploits the connection between kernel smoothing and diffusion, can be another option. However, we observe that the heat kernel is not appropriate for our application because the computation time increases quadratically with increasing bandwidth. For practical implementation, we use the convolution kernel estimator \citep{rakshit2019fast}. This convolution kernel estimator of the intensity with Jones-Diggle correction is given by
\begin{align}
K_S(\mathbf{x}_i; \mathbf{c}_j^s, w_{s}) =  \frac{\kappa(\mathbf{x}_i; \mathbf{c}_j^s,w_s)}{c_{\mathcal{L}}(\mathbf{c}_j^s)},
\label{eq:kappa}
\end{align}
where $c_\mathcal{L}(\mathbf{c}_j^s) = \int_\mathcal{L} \kappa(\boldsymbol{v};\mathbf{c}_j^s,w_s)d_1\boldsymbol{v}$ is a correction term, and $\boldsymbol{v}$ is an arbitrary point on $\mathcal{L}$. In \eqref{eq:kappa}, the numerator $\kappa$ is a planar Gaussian kernel with a smoothing bandwidth $w_s$. Note that we can interpret it as a spatial range of clusters since the bandwidth is equivalent to the standard deviation of the Gaussian kernel \citep{chiu1991bandwidth}. The denominator in \eqref{eq:kappa} is a convolution of the kernel $\kappa(\cdot)$ with the arc-length measure on the network, and thus the kernel $K_S(\cdot)$ can account for the geometrical structure of the street network. In particular, the correction term $c_{\mathcal{L}}(\mathbf{c}_j^s)$ can be obtained through Monte Carlo integration
\begin{equation}
c_\mathcal{L}(\mathbf{c}_j^s) \approx \cfrac{|\mathcal{L}|}{N} \sum_{k=1}^{N} \kappa(\boldsymbol{v}_k; \mathbf{c}_j^s, w_s), 
\label{MCapproximation}
\end{equation}
where $\boldsymbol{v}_1,\cdots,\boldsymbol{v}_N$ are generated from $\mathcal{L}$, which is a uniform distribution over the linear network. In our case, we observe that $N=1,000$ can provide an accurate Monte Carlo approximation. 

We thus model the membership variable $g_i \sim \text{Categorical}(\tilde{q}_{i1}, \ldots, \tilde{q}_{iM})$, where each $\tilde{q}_{ij}$ is set as 
\begin{align}
    \tilde{q}_{ij} &\propto q_j 
    K_S(\mathbf{x}_i;\mathbf{c}_j^s, w_{s})K_T(t_i; c_{j}^{t}, w_{t}), \quad j = 1, \dots, M.
    \label{calculate_q_ij}
\end{align}
Equation \eqref{calculate_q_ij} implies that it is likely that the $i$th observation is assigned to the $j$th cluster if $(\mathbf{x}_i,t_i)$ is close to $(\mathbf{c}_j^s, c_{j}^{t})$. In \eqref{eq:stick-breaking}, we assign the stick-breaking prior for $q_j$, and this ensures only $M^{*} (<M)$ non-empty clusters. From the mixture model specified above, the full likelihood function is of the form
\begin{align}
L(\boldsymbol{\theta}|\mathbf{X},\mathbf{c}) \propto
& \prod_{i=1}^{N} \left[ \sum_{\substack{\forall j \text{non-empty}}}
q_{j} K_S(\mathbf{x}_i;\mathbf{c}_j^s, w_{s})K_T(t_i; c_{j}^{t}, w_{t}) \right], 
\label{full_likelihood}
\end{align}
where $\boldsymbol{\theta} = (w_{s}, w_{t})$ is a model parameter. With a prior specification, the joint posterior distribution can be formulated as
\(\pi(\boldsymbol{\theta}, \mathbf{c}, \{ g_i \}_{i=1}^N, \{ U_j \}_{j=1}^M, b_u | \mathbf{X})\). For $w_s$ and $w_t$, we use flat truncated normal priors to ensure the positivity of the variances. Since the average length of each street in the entire Valencia is around 92.65$m$ and points over 2$km$ apart are considered distant for clustering, we set the lower and upper bounds for $w_s$ as 100 and 1,000 metres, respectively. Here, we also consider the criminological literature, which widely uses a 150$m$ $\times$ 150$m$ grid as a spatial unit \citep{block2000gang, santitissadeekorn2018}. This unit size reflects the largest area a single police officer typically covers on foot patrol within a designated beat \citep{zhang2022iml}. For $w_t$, we set the lower bound of the prior as 0. For \(\mathbf{c}\), we use the uniform prior defined over the network and the time period. 
%To accelerate computation, we discretize the network domain $\mathcal{L}$ through $50\times 50$ pixels $\Delta$. Then, we propose a spatial location of the cluster center from the discretized uniform prior.
To accelerate computation, we propose a spatial location of the cluster center uniformly from the set of line segments $\Delta$ longer than 500$m$. The hierarchical model structure and full conditionals for all components are described in the supplementary material. Algorithm \ref{alg:alg1} summarizes the Markov chain Monte Carlo (MCMC) procedure for Bayesian inference. 

\algrenewcommand\alglinenumber[1]{}
\begin{algorithm}[htbp]
\caption{MCMC algorithm for an STDP on a linear network}
\label{alg:alg1}
    \begin{algorithmic}
        \State Given \(\left\{\boldsymbol{\theta}^{(b)}, \mathbf{c}^{(b)}, \{ g_i^{(b)} \}_{i=1}^N, \{ U_j^{(b)} \}_{j=1}^M, b_u^{(b)}\right\}\) at the $b$-th iteration
        \vspace{0.5cm}
        
        \State \textbf{Update the model parameters} $\boldsymbol{\theta}^{(b+1)}:$\\
        Propose \(\boldsymbol{\theta}^* \sim q(\cdot|\boldsymbol{\theta}^{(b)})\) and accept \(\boldsymbol{\theta}^{(b+1)} = \boldsymbol{\theta}^*\) with probability
        \[
        \alpha = \min \left\{1,
        \frac{\pi\left(\boldsymbol{\theta}_{}^*|\mathbf{c}^{(b)}, \{g_i^{(b)}\}_{i=1}^N, \{U_j^{(b)}\}_{j=1}^M, b_u^{(b)}\right)}
        {\pi\left(\boldsymbol{\theta}_{}^{(b)}|\mathbf{c}^{(b)}, \{g_i^{(b)}\}_{i=1}^N, \{U_j^{(b)}\}_{j=1}^M, b_u^{(b)}\right)}\right\}
        \]
        \vspace{0.5cm}
        
        \State \textbf{Update cluster index probability for each observation}:\\
        Generate \(g_i^{(b+1)} \sim \text{Categorical}(\tilde{q}_{i1}, \ldots, \tilde{q}_{iM})\), where 
        \[\tilde q_{ij} \propto q_j K_S(\mathbf{x}_i;\mathbf{c}_j^s, w_{s})K_T(t_i; c_{j}^{t}, w_{t})\]
        for \(i = 1, \ldots, N\) and \(j = 1, \ldots, M\).
        \vspace{0.5cm}
        
        \State \textbf{Update Dirichlet process parameters}:\\
        Generate
        \[U_j^{(b+1)} \sim \text{Beta}(1 + \sum_{i=1}^N \mathbbm{1}\{g_i^{(b+1)} = j\}, b_u^{(b)} + \sum_{i=1}^N \mathbbm{1}\{g_i^{(b+1)} > j)\}\]
        for \(j = 1, \cdots, M\).\\
        \vspace{0.5cm}
        Generate
        \[ b_u^{(t+1)} \sim \text{Gamma}(M, c - \sum_{j=1}^{M-1} \log(1 - U_j^{(t+1)}))\]
        \vspace{0.5cm}  
        
        \State \textbf{Update space-time cluster centers} \(\mathbf{c}^{(b+1)}\):\\
        Propose \((\mathbf{c}_j^{s*}, c_j^{t*}) \sim \cfrac{1}{|\Delta|\mathcal{T}}\) and accept \((\mathbf{c}_j^{s(b+1)}, c_j^{t(b+1)}) = (\mathbf{c}_j^{s*}, c_j^{t*})\) with probability
        \[
        \alpha = \min \left\{1,
        \frac{\pi\left((\mathbf{c}_j^{s*}, c_j^{t*})|\boldsymbol{\theta}^{(b+1)}, \{g_i^{(b+1)}\}_{i=1}^N, \{U_j^{(b+1)}\}_{j=1}^M, b_u^{(b+1)}\right)}
        {\pi\left((\mathbf{c}_j^{s(b)}, c_j^{t(b)})|\boldsymbol{\theta}^{(b+1)}, \{g_i^{(b+1)}\}_{i=1}^N, \{U_j^{(b+1)}\}_{j=1}^M, b_u^{(b+1)}\right)} \right\}
        \]
        for \(j = 1, \ldots, M\).
    \end{algorithmic}
\end{algorithm}

\subsection{Post-Processing}

Post-processing is required to obtain the final results from DP-based methods due to the label switching issue \citep[see, \textit{e.g.},][]{medvedovic2002bayesian,dahl2006model,reich2011spatial}, where labels for any two different clusters can be swapping at a certain point of the MCMC chain. Without post-processing, each cluster label may represent multiple clusters in the data, leading to spurious multi-modality and overestimated variability. To address this, we identify the $b^*$th iteration of the MCMC chain that minimizes a loss related to cluster memberships, as follows:
\[
b^*=\arg \min_{b}
\left\{
\sum_{i=1}^N\sum_{j=1}^N\{\mathbbm{1}{ (g_{i}^b = g_{j}^b)-d_{ij} } \}^2
\right \}, \quad d_{ij} = \sum_{b=1}^B\{ \mathbbm{1}{ (g_{i}^b = g_{j}^b) }/B \}
\]
Here, \textit{B} denotes the total number of iterations in the MCMC, and $g_i^b$ denotes the membership of the sampled segment for the \textit{i}th observation in the \textit{b}th iteration. This solution minimizes the posterior expected loss for estimated segment membership proposed by \cite{binder1978bayesian} in Bayesian model-based clustering \citep{dahl2006model}.

\subsection{Model assessment}

We evaluate the performance of our model by comparing the empirical proportion of data points with the theoretical proportion derived from the likelihood function. Consider the rectangular cubes covering our spatio-temporal domain. The theoretical proportion corresponding to the cube  \([\alpha_{g},\beta_{g}]^3\) is calculated as
\[
p_{g}^t = \int_{\substack{[\alpha_{g},\beta_{g}]^3 \in \mathcal{L} \times \mathcal{T}}} \sum_{\forall j \text{ non-empty}} q_j K_S(\mathbf{x}; \mathbf{c}_j^s, w_{s})K_T(t; c_{j}^{t}, w_{t}) d\textbf{X},
\]
%\label{eq:modelassessment}
where \(\textbf{X} \in \mathcal{L} \times \mathcal{T}\) represents the spatio-temporal point process. Since the integral in the above equation is analytically intractable, we compute an approximation, \(\hat{p}_{g}^t\), through numerical integration. This is implemented with the \texttt{pmvnorm} function from the \texttt{R} package \texttt{mvtnorm}. Similarly, we compute the observed proportion of data points corresponding to the cube \([\alpha_{g},\beta_{g}]^3\) as
\[
p_{g}^o = \cfrac{\text{the number of observations in \([\alpha_{g},\beta_{g}]^3\)}}{\text{total number of observations}}.
\]
We compare $\hat{p}_{g}^t$ and $p_{g}^o$ computed on each cube. Note that if we use too small cubes, we would have a lot of zeros for $p_{g}^o$ because there will be no observation in \([\alpha_{g},\beta_{g}]^3\). On the other hand, if we use too large cubes, the approximation error for $\hat{p}_{g}^t$ becomes larger because the cubes will also cover the region without the roads. To address this, we first construct $G_s = 200 \times 200 \times 10$ number of subgrid points (i.e., high-resolution grid) over \(\mathcal{L} \times \mathcal{T}\) and compute $\hat{p}_{g}^t$ and $p_{g}^o$ at each cube. Then, we aggregate them into larger, nonoverlapping grid points $G = 5 \times 5 \times 10$ that also cover $\mathcal{L} \times \mathcal{T}$. We provide details on grid computation in the supplementary material.

The proposed assessment method is related to residual analysis, which is commonly used for point process model validation \citep{baddeley2005residual}. This approach also computes the discrepancies between observed proportions $p_g^o$ and the estimated theoretical proportions $\hat{p}_g^t$ to generate residual maps; residual values distributed around zero indicate a good model fit. Note that $p_g^o$ and $\hat{p}_g^t$ are the same quantities defined earlier in this section. We present the results in the supplementary material, where our model shows better agreement with the observed data compared to the standard DP model \citep{park2023spatio}.

%%%%%%%%%%%%%%%%%%%%%%%%%%%%%%%%%%%%%
\section{Application} \label{Application}

We applied our proposed method to the Valencia crime data described in Section \ref{data_description}. The code was implemented in \texttt{R} version 4.1.3 \citep{R} and executed on an AMD Ryzen 9 5950X 16-core processor.
The complete MCMC analysis, comprising 20,000 iterations and post-processing, required approximately 23 hours for crime type robbery,
and 9 hours for crime type theft. The maximum number of clusters was set to 80 for robbery and
70 for theft, with the number of detected clusters automatically determined to be below these
limits. Here, note that we need to penalize the maximum number of cluster centers. Setting the maximum number of clusters excessively high with small range parameters results in dense centers throughout the region, leading to overfitting. However, if there are few centers, it may not detect centers in the area.

\subsection{Crime hotspots}
    
\begin{figure}[htbp]
        \centering
                \centering
                \includegraphics[width=0.95\linewidth]{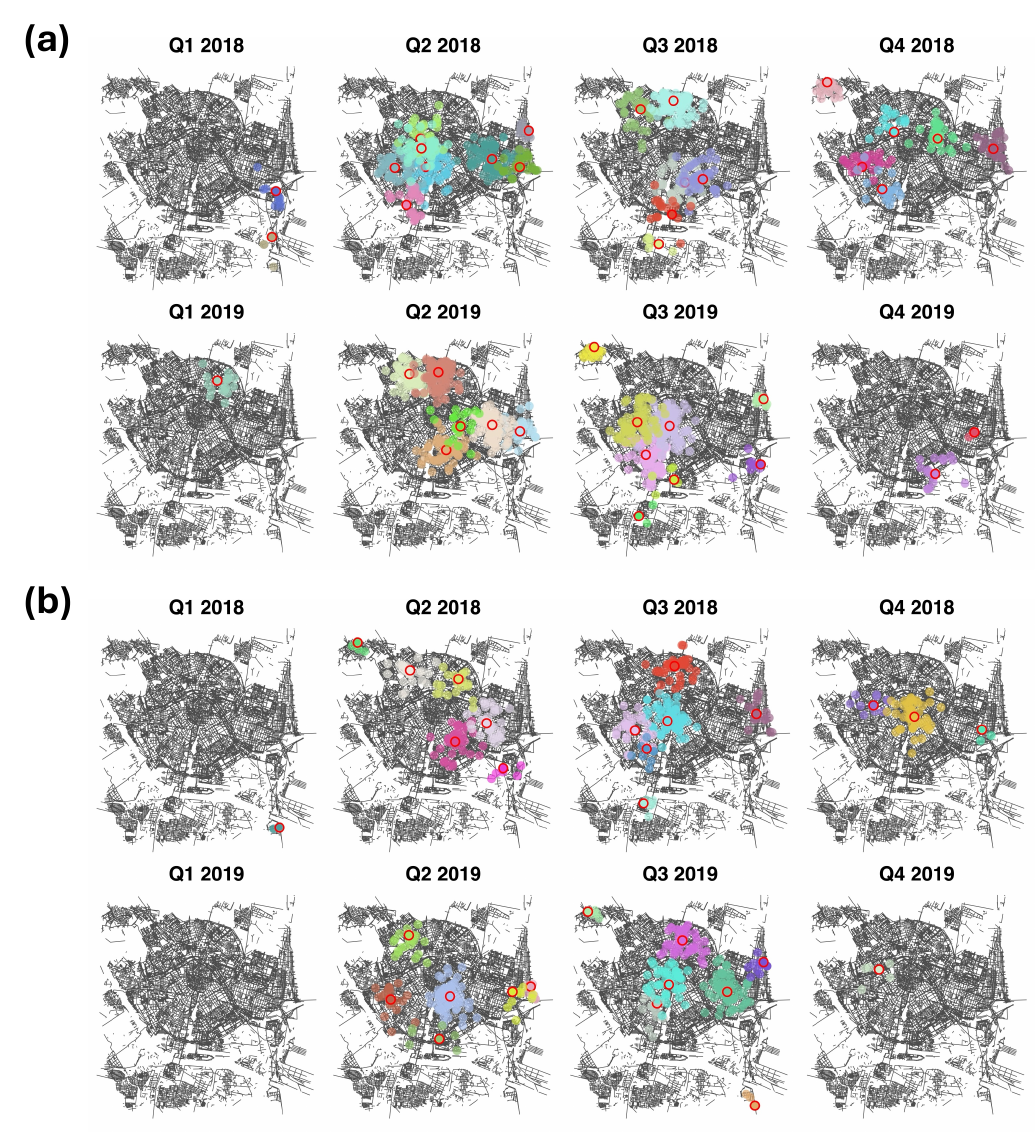}
        \caption{Detected cluster centers and their offspring by quarters of (a) robbery and (b) theft. The red-outlined dots indicate detected cluster centers, while the surrounding dots of the same color represent their offspring belonging to the same cluster.}
        \label{fig:detected_cc_off}
    \end{figure}

Figure~\ref{fig:detected_cc_off} shows detected cluster centers and their offspring for each crime type, analyzed quarterly over the two-year period. Robbery cluster centers were more frequently detected in Q2 of 2018 and Q3 of 2019, while theft clusters were more commonly detected in Q3 of 2019. These findings align with prior research on seasonal variations in crime levels \citep{cohn1990, farrell1994, mcdowall2011}. Numerous studies have demonstrated that both violent and property and personal crimes tend to peak during the summer or fall months \citep{deckard2023, haberman2018, hipp2004, kim2022}. This pattern is likely attributable to shifts in individuals’ routine activities, such as vacation and outdoor events, which increase the likelihood of convergence between potential offenders and suitable victims in the absence of effective guardianship \citep{cohen1979}. Additional evidence supporting this argument is the peak in tourist numbers in Valencia between May and October (Q2 and Q3) in both 2018 and 2019 \citep{ine2025}.
    % \begin{figure}[htbp]
    %     \centering
    %         \begin{minipage}{\linewidth}
    %             \centering
    %             \includegraphics[width=0.95\linewidth]{figure/agg cc and off.jpg}
    %             \subcaption{(a) Robbery}
    %             \label{fig:agg_cc_off}
    %         \end{minipage}%
    %         \vspace{0.3cm}
    %         \begin{minipage}{\linewidth}
    %             \centering
    %             \includegraphics[width=0.95\linewidth]{figure/theft cc and off.jpg}
    %             \subcaption{(b) Theft}
    %             \label{fig:theft_cc_off}
    %         \end{minipage}%
    %     \caption{Detected cluster centers and their offspring by quarters. The red-outlined dots indicate detected cluster centers, while the surrounding dots of the same color represent their offspring belonging to the same cluster.}
    %     \label{fig:detected_cc_off}
    % \end{figure}

\begin{figure}[htbp]
    \centering
            \centering
            \includegraphics[width=\linewidth]{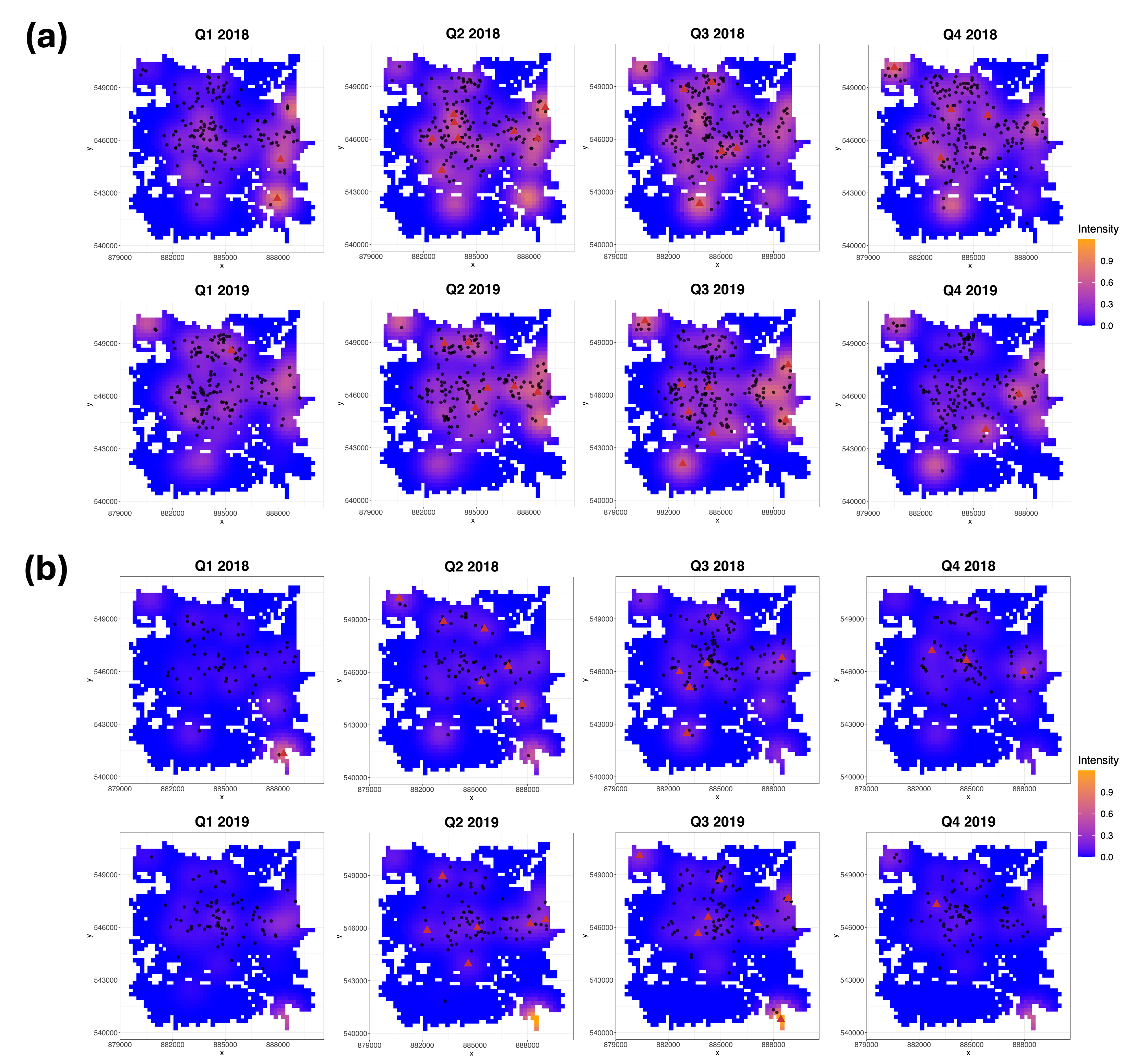}
    \caption{Posterior means of the density of the observed cases of (a) robbery and (b) theft by quarter. Black dots represent the observed crime events during each period, while red triangles indicate the detected cluster centers.}
    \label{fig:post_mean_dens}
    \end{figure}

Figure~\ref{fig:post_mean_dens} represents the posterior mean of the density on the linear network, calculated by our model \eqref{full_likelihood}, with the estimated range parameter plugged in. We observe that areas with higher observed counts correspond well to regions of higher posterior means, indicating a good model fit.
% This figure shows robbery highlighted in yellow in the upper area in Q2 of 2018 and in the bottom area in Q3 of 2019. For theft, the number of detected cluster centers is smaller than that for robbery, with a denser concentration in Q3 of 2018. Interestingly, the patterns of robbery and theft exhibit similarities in Q3 of 2018.
This figure shows robbery highlighted in yellow in the upper area in Q2 of 2018 and the bottom area in Q3 of both years. For theft, the number of detected cluster centers is smaller than that for robbery, with a denser concentration in Q3 of both years. Interestingly, the patterns of robbery and theft exhibit similarities in the right side in Q2 and Q3 of 2019. For both crime types, hotspots appear to be shifting from one quarter to the next. This observation supports the literature indicating that hotspots exhibit short-term stability, typically lasting only a few months \citep{deckard2023, groff2015}. Consequently, focusing police and social service resources in crime hotspots for brief periods may be effective in achieving notable reductions in crime \citep{weisburd2018}.

    % \begin{figure}[htbp]
    % \centering
    %     \begin{minipage}{\linewidth}
    %         \centering
    %         \includegraphics[width=\linewidth]{figure/agg int.jpg}
    %         \subcaption{(a) Robbery}
    %         \label{fig:spatial intensity_aggression}
    %     \end{minipage}%
    %     \vspace{0.2cm}
    %     \begin{minipage}{\linewidth}
    %         \centering
    %         \includegraphics[width=\linewidth]{figure/theft int.jpg}
    %         \subcaption{(b) Theft}
    %         \label{fig:spatial intensity_theft}
    %     \end{minipage}%
    % \caption{Posterior means of the density of the observed cases of robbery and theft by quarter. Red triangles indicate the detected cluster centers during each period.}
    % \label{fig:post_mean_dens}
    % \end{figure}

    % \begin{figure}[h]
    %     \centering
    %     \begin{minipage}[t]{0.65\linewidth}
    %         \centering
    %         \includegraphics[width=\linewidth]{figure/scatterplot.jpg}
    %         \subcaption{(a)}
    %         \label{fig:scatterplot}
    %     \end{minipage}%
    %     \hfill
    %     \begin{minipage}[t]{0.34\linewidth}
    %         \centering
    %         \includegraphics[width=\linewidth]{figure/multipcf.jpg}
    %         \subcaption{(b)}
    %         \label{fig:multipcf}
    %     \end{minipage}
    %     \caption{(a): Scatter plots of observed proportion and theoretical proportion (Left: robbery, Right: theft), (b): Multitype pair correlation function. The black solid line represents the inhomogeneous estimate of the pair correlation function and the red dashed line represents 1, the uniform Poisson process.}
    %     \label{fig:scatterandpcf}
    % \end{figure}

\begin{figure}[htbp]
    \centering
    \includegraphics[width=\linewidth]{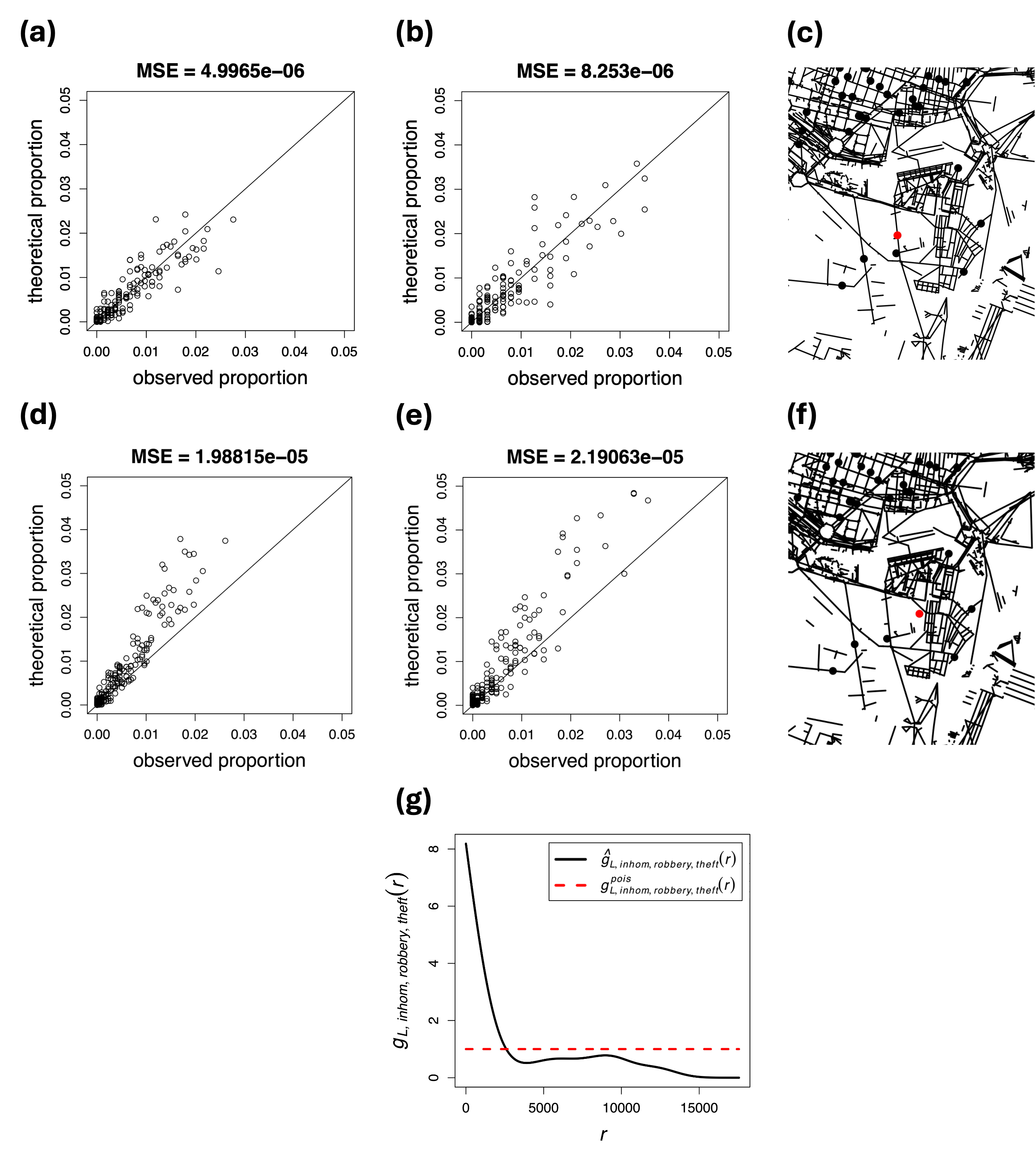}
    \caption{Model fitting comparison between the linear network DP model and the standard DP model. (a), (b) Scatter plots of $(p_g^o, \hat{p}_g^t)$ for robbery and theft under the linear network model; (d), (e) results from the standard DP model. (c), (f) Cluster centers obtained from the linear network and standard models, respectively. Black dots represent observed crime events, and red dots indicate cluster centers. (g) Multitype pair correlation function. The black solid line represents the inhomogeneous estimate of the pair correlation function and the red dashed line represents 1, the uniform Poisson process.}
\label{fig:scatter_cc}
\end{figure}
    % \begin{figure}[h]
    %     \centering
    %         \centering
    %         \includegraphics[width=\linewidth]{figure/figure6.pdf}
    %     \caption{(a): Scatter plots of observed proportion and theoretical proportion (Left: robbery, Right: theft), (b): Multitype pair correlation function. The black solid line represents the inhomogeneous estimate of the pair correlation function and the red dashed line represents 1, the uniform Poisson process.}
    %     \label{fig:scatterandpcf}
    % \end{figure}
% Figure~\ref{fig:scatterandpcf}a shows scatter plots of the estimated and observed proportions, \(\hat{p}_{g}^t\) and \(p_{g}^o\). The points are scattered around the straight one-to-one line for both crime types, indicating that our model effectively captures the spatio-temporal densities of points on a linear network.
To assess model fitting, we compare our proposed model with the standard spatio-temporal DP model \citep{park2023spatio}, which does not account for the network structure. The standard DP model shares the same structure as our linear network DP model but uses a planar Gaussian kernel for $K_S(\cdot)$ in \eqref{eq:spatialkernel}, thereby ignoring the geometry of the underlying road network. Figure~\ref{fig:scatter_cc} shows scatter plots of the estimated and observed proportions, $\hat{p}^t_g$ and $p^o_g$, from both models. We observe that, for both crime types, the points from our model lie closer to the one-to-one line compared to those from the standard DP model, and the mean squared error (MSE) is also smaller. This indicates that our model more effectively captures the spatio-temporal point densities on a linear network. The limitations of the standard DP model in accounting for the road structure are also evident in the detected cluster centers. As shown in Figure~\ref{fig:scatter_cc}c, our linear network model identifies clusters along the road network, whereas the standard DP model detects cluster centers in the continuous planar domain (Figure~\ref{fig:scatter_cc}f). This demonstrates that our model yields more realistic and interpretable cluster locations.
We fitted the model given by Equation (\ref{full_likelihood}) to the data described in Section \ref{Data} using Algorithm \ref{alg:alg1}. The estimated parameters are summarized in Table \ref{tab:posterior_w}. 
% For robbery, the space and time range parameters are about 545$m$ and 0.14 (roughly 3 months and 11 days) respectively, indicating that 95\% of observed points lie within approximately circles of 2$\times$545$m$ and temporal intervals of 6 months and 22 days.
% For theft, the space and time range parameters are about 563$m$ and 0.142 (roughly 3 months and 12 days). Accordingly, 95\% of observed points fall within circles of 2$\times$563$m$ and within temporal intervals of 6 months and 24 days. With this information at hand, we can delineate the spatio-temporal structure and behavior of our crime data. These estimates align with the observation that both types of crime exhibit temporal concentration in two seasonal quarters—Q2 and Q3, which correspond to the summer and fall months. Additionally, the findings suggest that the spatial diameter of each crime cluster is approximately 1.1 kilometers, with theft clusters being slightly more dispersed than robbery clusters.

% \begin{table}[htbp]
% \centering
% \begin{tabular}{ccc}
% \hline
%   & $w_s$ (space) & $w_t$ (time) \\
%   \hline
% Robbery &  544.798 & 0.140 \\
%      &  (511.308, 572.985) & (0.131, 0.150)\\
% Theft &  563.402 & 0.142\\
%      &  (536.571, 587.816) & (0.131, 0.155) \\
% \hline
% \end{tabular}
% \caption{\textcolor{red}{Posterior means and 95\% HPD intervals (parentheses) of space and time range parameters for robbery and theft.}}
% \label{tab:posterior_w}
% \end{table}
For robbery, the space and time range parameters are about 569$m$ and 0.149 (roughly 3 months and 18 days) respectively, indicating that 95\% of observed points lie within approximately circles of 2$\times$569$m$ and temporal intervals of 7 months.
For theft, the space and time range parameters are about 552$m$ and 0.141 (roughly 3 months and 12 days). Accordingly, 95\% of observed points fall within circles of 2$\times$552$m$ and within temporal intervals of 6 months and 24 days. With this information at hand, we can delineate the spatio-temporal structure and behavior of our crime data. These estimates align with the observation that both types of crime exhibit temporal concentration in two seasonal quarters—Q2 and Q3, which correspond to the summer and fall months. Additionally, the findings suggest that the spatial diameter of each crime cluster is approximately 1.1 kilometers, with robbery clusters being slightly more dispersed than theft clusters.

\begin{table}[htbp]
\centering
\begin{tabular}{ccc}
\hline
  & $w_s$ (space) & $w_t$ (time) \\
  \hline
Robbery &  569.322 & 0.149 \\
     &  (538.078, 596.802) & (0.142, 0.158)\\
Theft &  551.732 & 0.141\\
     &  (523.050, 583.479) & (0.129, 0.154) \\
\hline
\end{tabular}
\caption{Posterior means and 95\% HPD intervals (parentheses) of space and time range parameters for robbery and theft.}
\label{tab:posterior_w}
\end{table}

\subsection{Visualization of risk boundaries} 
\begin{figure}[htbp]
	\centering
	\includegraphics[width=\linewidth]{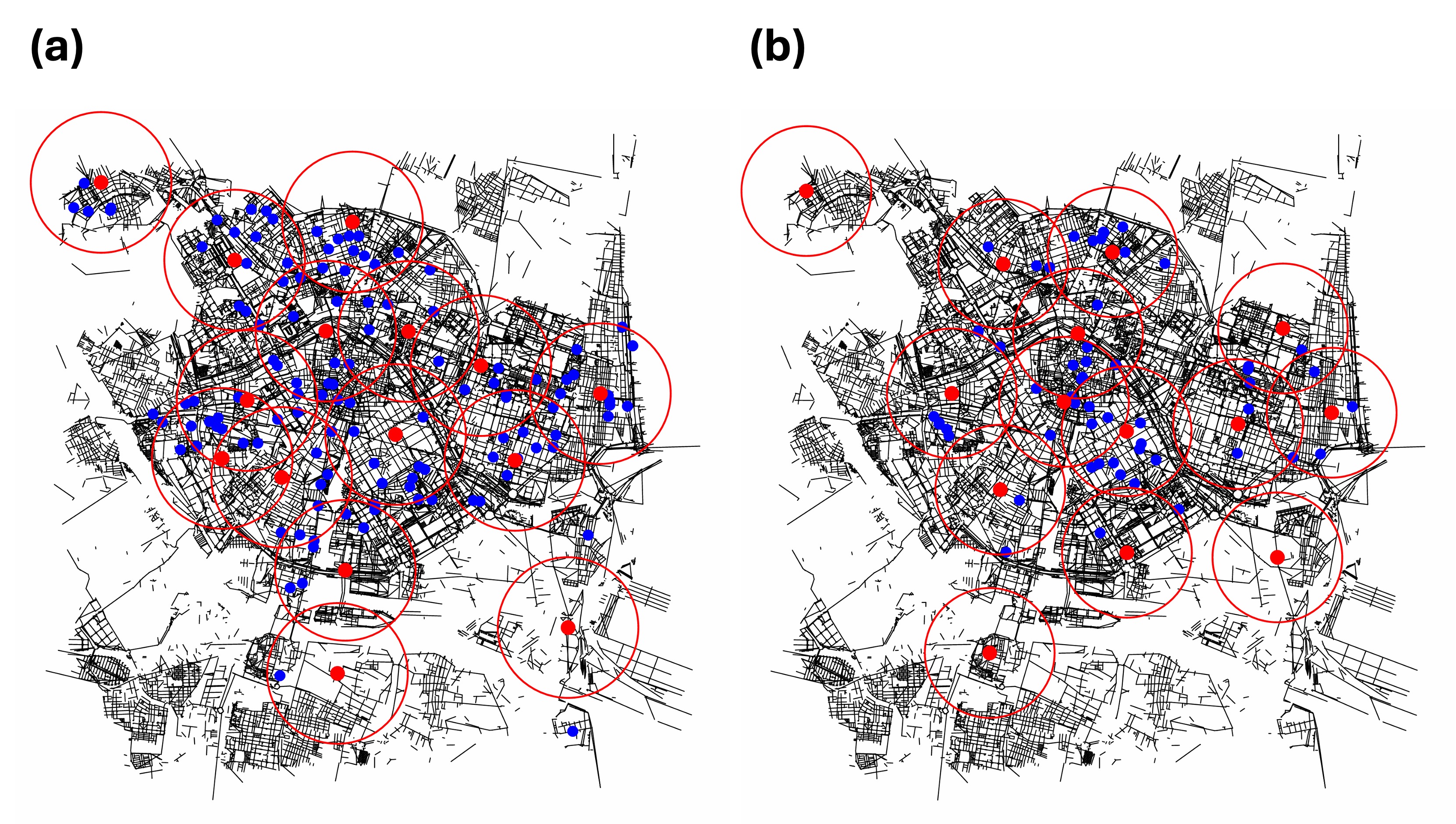}
	\caption{Risk boundaries estimated from a five-month period (2018-05-15 to 2018-09-27) for (a) robbery and (b) theft. Red circles represent risk boundaries, and red dots denote cluster centers. Blue dots indicate observed crime incidents during the subsequent two-month period (2018-10-01 to 2018-11-30).
    }
\label{fig:predict}	
\end{figure}

Our model is designed as an unsupervised learning approach, with the primary objective of describing spatio-temporal point patterns from data, rather than for prediction. However, based on the estimated values of $w_s$ and $w_t$ from past data, it is possible to construct "spatio-temporal risk boundaries", which can be used to identify areas with a high risk of crime events over short time periods—assuming that the event distribution during such periods can be approximated by a model fitted to historical data. Such boundaries may be useful for informing social interventions to prevent crime incidents. Here, we fit the model using data from a five-month period (2018-05-15 to 2018-09-27). For robbery, the posterior means of $w_s$ and $w_t$ are estimated to be 567.889$m$ and 0.05 (approximately 5 weeks), respectively. For theft, the posterior means of $w_s$ and $w_t$ are 524.141$m$ and 0.047 (approximately 1 month), respectively. Figure~\ref{fig:predict} visualizes the cluster centers, their spatial risk boundaries represented by circles with radius $2\times w_s$, and future crime events observed over the subsequent two-month period (2018-10-01 to 2018-11-30), corresponding to $2\times w_t$. Most of the crime events fall within the risk boundaries, indicating that the model can identify areas with a high risk of crime events.

\subsection{Dependence of crime incidence on amenity locations}
\begin{figure}[htbp]
        \centering
        \includegraphics[width=\linewidth]{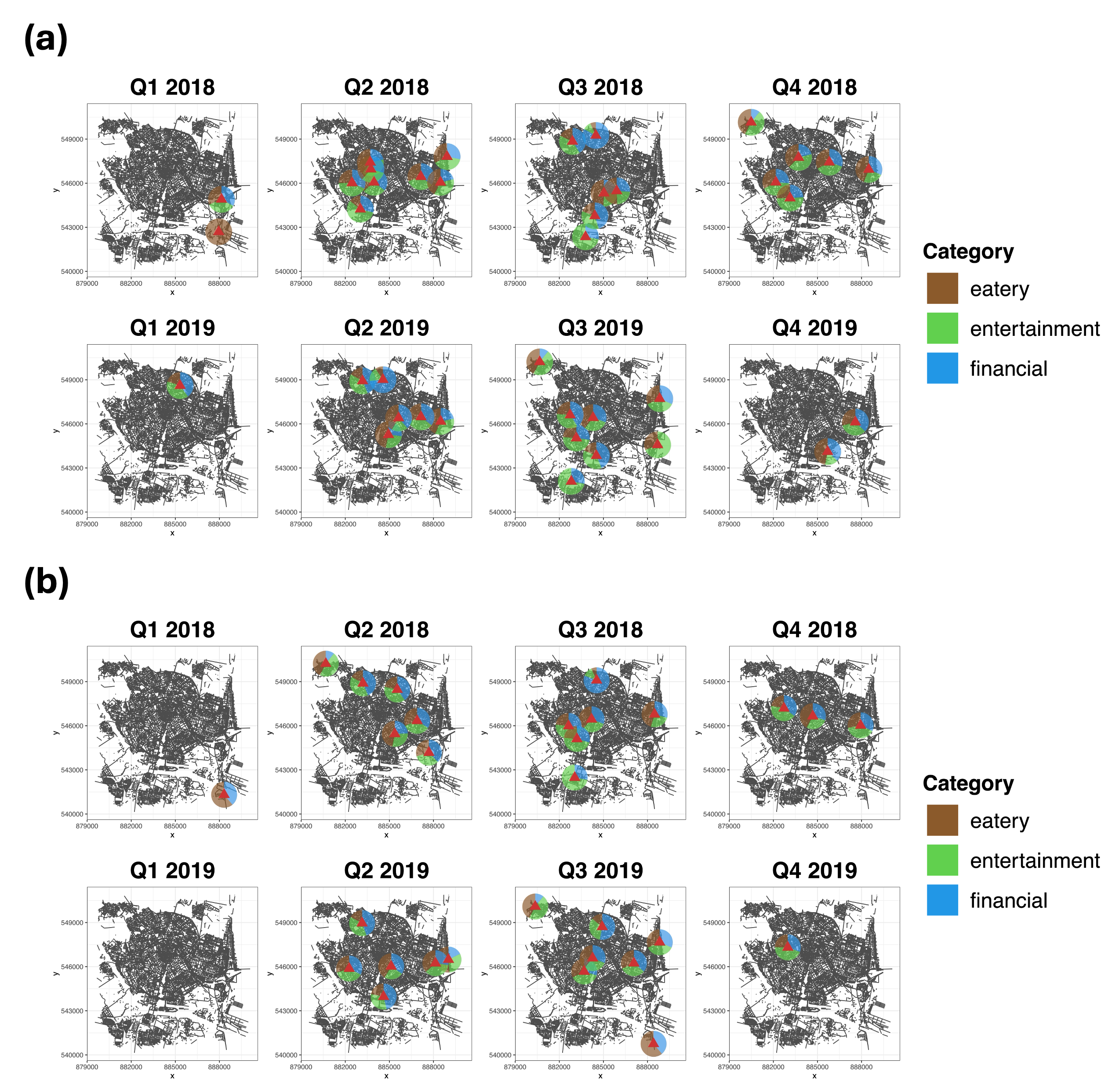}
        \caption{Amenity proportions for each cluster by quarter. Red triangles indicate detected cluster centers. The brown, green, and blue sections show proportions of eatery, entertainment, and financial amenities, respectively. (a) robbery and (b) theft.}
        \label{fig:int_am_pie}
    \end{figure}

Figure~\ref{fig:int_am_pie} illustrates the relative proportions (score) of different amenity types located within the spatial risk boundaries (i.e., $2\times \omega_s$) from each cluster center. A higher proportion of a specific amenity within a cluster’s boundary suggests a potential association with the occurrence of crime events. To adjust for the overall frequency of each amenity type, we compute the following normalized score:
\[
\text{Score}_{k}^{(j)} = \# \text{amenity}_k \text{ in cluster } j \times \left( \frac{\# \text{total amenities}}{\# \text{amenity}_k \text{ in entire area}} \right),
\]
where a higher score indicates a local concentration of amenity type $k$ in cluster $j$ relative to its global frequency. The relative proportion of amenity types varies across clusters. Overall, the distribution of amenities is uneven, with the city center exhibiting a higher concentration of financial and entertainment facilities, whereas the outskirts predominantly feature eateries. 

According to routine activity theory \citep{cohen1979}, crime is most likely to occur where motivated offenders, suitable targets, and the absence of capable guardians converge. When integrated with research on crime specificity, which emphasizes that “even subtypes of a general crime category [may] have different causes, require different situational contexts, occur at different times and in different places, and are likely prevented by different methods” \citep[p. 432]{Haberman2022}, the findings suggest that robberies in city centers are more likely to be driven by opportunities associated with financial and entertainment facilities (e.g., banks, ATMs, bars, nightclubs), whereas robberies occurring on the outskirts may be more closely linked to eateries (e.g., cafes, restaurants).

% Figure~\ref{fig:int_am_pie} illustrates the relative proportions of amenities located within the 2 $\times$ (spatial range) from each cluster center.
% To prevent bias, the weight is calculated as the ratio of the total number of amenities to the count of a specific amenity within the buffer, and the final value is calculated by multiplying this ratio by the specific amenity count.
% The spatial closeness of amenities to cluster centers suggests their potential association with the occurrence of crime events. The proportion of amenity types varies across clusters, and a high proportion of a specific amenity type indicates that these amenities are more influential in the cluster. Overall, the distribution of amenities is uneven, with city center exhibiting a higher concentration of financial and entertainment facilities, whereas the outskirts predominantly feature eateries and some financial services (see again Figure~\ref{fig:int_am_pie}).

Clusters situated in the bottom-right quadrant, a coastal area, indicate that both crime categories are exclusively influenced by eateries. Note that this area of the city sits around the coast and a lake (called \textit{Albufera}) and is full of restaurants and open-air dinning places. Thus, our model offers additional evidence that eateries generate opportunities for crime due to high-traffic nature, elevated levels of activity and visibility, and the presence of stored cash \citep{brantingham1995}. The bottom central area, corresponding to the downtown district, is characterized by a high concentration of entertainment-related amenities that attract pedestrians for recreational purposes, while the top central area is primarily defined by the presence of financial establishments. 

Throughout the study period, entertainment venues and eateries were more prominently represented in crime clusters, compared to financial amenities. These findings align with prior research examining the relationship between land use and crime (for a review, see \citep{wilcox2011}). Specifically, commercial land use is associated with higher crime rates, residential land use with lower crime rates, and office and industrial land uses with increased property crime \citep{boessen2015, haberman2015, kinney2008, lee2022, stucky2009}. However, this study only examined amenities related to commercial land use, excluding other types of land use.

    % \begin{figure}[htbp]
    %     \centering
    %         \begin{minipage}{\linewidth}
    %             \centering
    %             \includegraphics[width=\linewidth]{figure/agg int am pie.jpg}
    %             \subcaption{(a) Robbery}
    %             \label{fig:agg_int_am_pie}
    %         \end{minipage}%
    %         \vspace{0.2cm}
    %         \begin{minipage}{\linewidth}
    %             \centering
    %             \includegraphics[width=\linewidth]{figure/theft int am pie.jpg}
    %             \subcaption{(b) Theft}
    %             \label{fig:theft_int_am_pie}
    %         \end{minipage}%
    %     \caption{Amenity proportions for each cluster by quarter. Red triangles indicate detected cluster centers. The brown, green, and blue sections show proportions of eatery, entertainment, and financial amenities, respectively.}
    %     \label{fig:int_am_pie}
    % \end{figure}

\begin{figure}[htbp]
	\centering
	\includegraphics[scale=0.38]{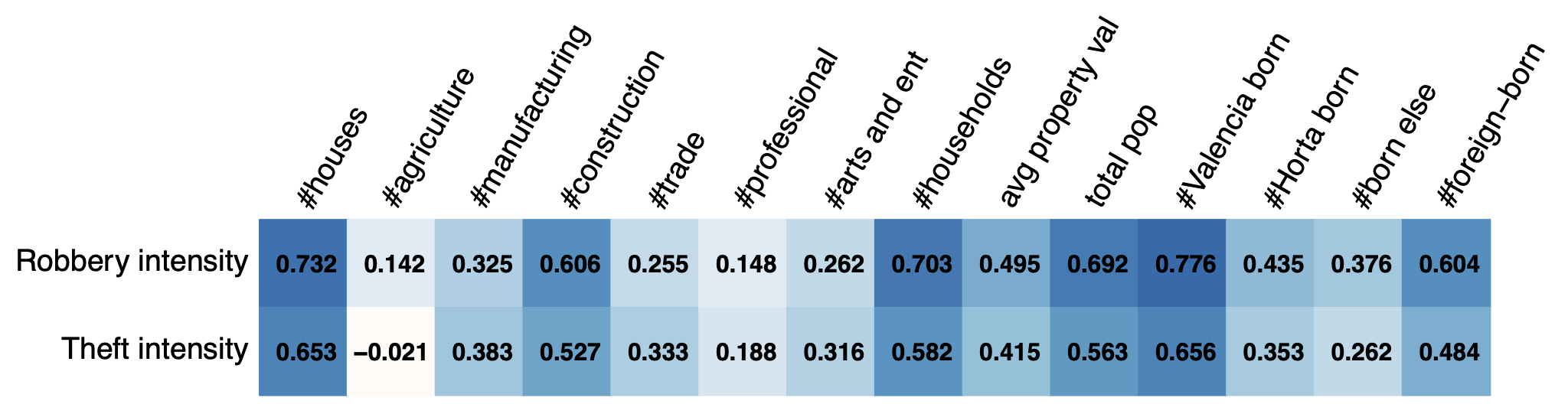}
	\caption{Pearson correlations between crime intensities and socio-economic variables. Blue colors indicate positive correlations, with darker shades representing stronger correlation magnitudes.}
\label{fig:corr}
\end{figure}

We further explored the relationships between socio-economic variables and crime intensity in Valencia using Pearson correlations. We collected 14 socio-economic variables for the years 2018 and 2019 across 19 districts in Valencia. These variables include: the number of housing units; the number of employment in agriculture, manufacturing, construction, retail and service sectors, professional occupations, arts and entertainment; the number of households; the average real estate price per square meter; the total population; the number of residents born in Valencia; the number of residents born in the Horta region; the number of residents born in other regions of Spain; and the number of foreign-born residents. We then computed the average value of each variable within each district. In addition, for each district, we computed the aggregated crime intensities for both crime types over the entire study period (2018–2019). Figure~\ref{fig:corr} illustrates that the crimes examined in this study tend to be concentrated in urbanized, densely populated districts, and occur much less frequently in predominantly rural areas. Crimes that occurred in Valencia exhibit a strong positive correlation with the number of Valencia-born residents. In contrast, the numbers of residents born in Horta and other regions of Spain show relatively weaker correlations compared to other population groups. The Horta district is marked by a relatively stable population, with most residents having been born there and few newcomers relocating for work or residence. These findings are consistent with criminological theories \citep{shaw1942, sampson1997}, which posit that crime rates tend to be higher in communities characterized by urbanization and residential instability.

\subsection{Association between the two crime types}

To analyze the association between the two crime types, we conduct a descriptive analysis using an inhomogeneous multitype pair correlation function (pcf) on a linear network \citep{baddeley2014multitype}. Specifically, after obtaining the spatial cluster centers for robbery and theft by fitting the linear DP model, we plug these centers into the inhomogeneous multitype pair correlation function to quantify how the two types of cluster centers co-locate. The inhomogeneous multitype pair correlation function is estimated as
\begin{equation}
\hat{\rho}^{\mathcal{L}, \text{inh}}(r) = \frac{1}{|\mathcal{L}|} \sum_{\mathbf{c}^{s}_{1,i} \in \mathbf{C}_1} \sum_{\mathbf{c}^{s}_{2,j} \in \mathbf{C}_2}
\frac{\kappa(r;d_\mathcal{L}(\mathbf{c}^{s}_{1,i}, \mathbf{c}^{s}_{2,j}),h)} 
{\hat{\lambda}_1(\mathbf{c}^{s}_{1,i}) \hat{\lambda}_2(\mathbf{c}^{s}_{2,j}) M( \boldsymbol v, d_\mathcal{L}(\mathbf{c}^{s}_{1,i},\mathbf{c}^{s}_{2,j})) } ,
    \label{multipcf}
\end{equation}
where $\mathbf{C}_1$ and $\mathbf{C}_2$ denote the spatial point processes of cluster centers for robbery and theft on the linear network $\mathcal{L}$, and $\mathbf{c}^{s}_{1,i}$ and $\mathbf{c}^{s}_{2,j}$ represent their $i$th and $j$th realizations, respectively. This summary statistic is used purely as a descriptive exploratory tool to assess whether the two crime-type clusters exhibit attraction (pcf $>1$) or repulsion (pcf $<1$) at different spatial distances along the network. In \eqref{multipcf}, $M( \boldsymbol v, d_\mathcal{L}(\mathbf{c}^{s}_{1,i},\mathbf{c}^{s}_{2,j}))$
is the number of locations $\boldsymbol v \in \mathcal{L}$ that are $d_\mathcal{L}(\mathbf{c}^{s}_{1,i},\mathbf{c}^{s}_{2,j})$ away from $\mathbf{c}^{s}_{1,i}$ \citep{ang2012geometrically} in terms of the shortest-path distance. $\kappa(r;d_\mathcal{L}(\mathbf{c}^{s}_{1,i}, \mathbf{c}^{s}_{2,j}),h)$ denotes a planar Gaussian kernel centered with the center $d_\mathcal{L}(\mathbf{c}^{s}_{1,i}, \mathbf{c}^{s}_{2,j})$ and the bandwidth $h$, which is determined using a Scott's rule \citep{scott2015multivariate}. We use the convolution kernel estimator \citep{rakshit2019fast} to obtain $\hat \lambda_1(\cdot)$ and $\hat \lambda_2(\cdot)$. Specifically, for each crime type  $m \in \{1, 2\}$, corresponding to robbery and theft, the intensity at the arbitrary point $\boldsymbol{u} \in \mathcal{L}$ is 
\[
\hat \lambda_m(\boldsymbol{u}) = \sum_{\mathbf{c}^{s}_{m,l}\in \mathbf{C}_m} \cfrac{\kappa(\boldsymbol{u};\mathbf{c}^{s}_{m,l},w_m)}{c_\mathcal{L}(\boldsymbol{u})},
\]
where $\kappa$ is a planar Gaussian kernel with a smoothing bandwidth $w_m$ and $c_\mathcal{L}(\boldsymbol{u}) = \int_\mathcal{L}\kappa(\boldsymbol{v};\boldsymbol{u},w_m)d_1\boldsymbol{v}$ is a correction term to consider the network geometry. The bandwidth parameter was chosen according to Scott’s rule \citep{scott2015multivariate}.

We computed a multitype pairwise correlation function between spatial cluster centers from the two crime types by using the \texttt{linearpcfcross.inhom} function from the \texttt{spatstat.linnet} package. From Figure~\ref{fig:int_am_pie}, we observe that the centers of the two crimes (red triangles) are close in Q3 of 2018 and Q2, Q3 of 2019. Indeed, the pcf plot (see Figure~\ref{fig:scatter_cc}g) takes values exceeding 1 at certain spatial distances ($r<2500$), indicating an attraction tendency. In contrast, we observe that the centers of the two crimes have a repulsive pattern in Q4 of 2019 (Figure~\ref{fig:int_am_pie}), and the computed pcf values drop below 1 at around $r>2500$. These results indicate that robbery and theft crime clusters are more concentrated in the fall when both peak and become more dispersed in other seasons with lower crime activity.

To study the association between the two crime types, we also fit a spatio-temporal conditional autoregressive (CAR) model \citep{lee2021carbayesst} by regressing the log-transformed robbery intensity on the log-transformed theft intensity, where both intensities are obtained using the convolution kernel estimator \citep{rakshit2019fast}. For the spatio-temporal regression, we use a $64\times64$ spatial grid, where each pixel is associated with an estimated intensity value at 8 time points corresponding to quarterly intervals over a two-year period. We provide details of the spatio-temporal CAR model and its implementation in the supplementary material. We observe that the regression coefficient is significantly positive, indicating that theft has a positive impact on robbery. Furthermore, autocorrelation parameters are statistically significant, suggesting the presence of spatial and temporal autocorrelation in the relationship between the two crime types.

%%%%%%%%%%%%%%%%%%%%%%%%%%%%%%%%%%%%%%%%%%%%%
\section{Discussion} \label{Discussion}

In this paper, we have proposed a Bayesian spatio-temporal point process model on a linear network. With a DP prior, the model can detect unobserved space-time cluster centers of crime incidence over the street network. From the detected cluster centers, we study the association between amenities and crime events, which provides interesting criminological research findings. Furthermore, we have estimated spatio-temporal intensities that can describe the pattern of crime events in the city. We use the computationally efficient convolution kernel estimator \citep{rakshit2019fast} in our model to account for the geometrical structure of the street network. We have also assessed the performance of the model by comparing the theoretical and empirical densities and observed that our model fits well. To our knowledge, this is the first attempt to consider the Bayesian framework for the point process analysis on a linear network. 

There are several caveats to be further considered in future research. An excessive number of cluster centers may lead to overfitting, but this issue can be mitigated through appropriate prior specification, as implemented in our work. The hyperparameter in the stick‐breaking process can be set to discourage an overly large number of non-empty clusters \citep{reich2011spatial}. In addition, specifying an informative prior on the range parameter can discourage the model from producing cluster centers with too small ranges \citep{park2023spatio}. Alternatively, the model could be extended to a supervised learning framework by splitting the data into training and test sets to improve out-of-sample generalization. In this study, we fitted the model to two crime types separately and analyzed the correlation between their cluster centers. Our model can be extended to a bivariate point process setting. Recently, \cite{briz2023mechanistic} proposed a mechanistic bivariate point process over the planar space, assuming that the current crime event is affected by different crime types in the past.
Another direction is to model the crime events as a marked point process that has information about the crime type. Several Bayesian approaches have been developed to model data with geographic locations and categorical marks \citep{jiao2021bayesian,quick2015bayesian}. 

Research on multitype marked point processes on linear networks is still in an early stage, although recent studies have made progress. For example, \cite{EckardtMoradi2024} reviewed and extended mark-related summaries to linear networks, highlighting that methodological developments for marked inhomogeneous processes on linear networks remain limited. Other contributions include extensions of summary characteristics to function-valued marks \citep{eckardt2025summary}, the use of compositional data analysis for marked spatial point processes \citep{eckardt2025spatial}, and the proposal of graph-based summaries such as the graph mark variogram to capture complex point patterns \citep{eckardt2024second}. These works mainly focus on summary characteristics rather than proposing new models. Incorporating such frameworks into our model would be an interesting extension.
There have been several proposals to study the relation between multiple types of events over time. For example, \cite{taddy2010autoregressive} developed an autoregressive mixture model with an autoregressive Beta stick-breaking process to study spatio-temporal relationships between extreme and non-extreme crimes. \cite{eckardt2021graphical} proposed partial point process characteristics to capture direct dependencies in multivariate marked spatio-temporal point processes. Extending such work to our application would require kernel estimators that account for network geometry, which would be computationally intensive. We therefore consider this direction as part of our future work.

\section*{Supplementary Material}
Supplementary materials available online contain full conditionals for the MCMC algorithm and grid computation details. 
 
\section*{Acknowledgement}
This work was supported by the National Research Foundation of Korea (RS-2025-00513129, RS-2025-00523567, RS-2023-00217705) and the New Faculty Startup Fund from Seoul National University (326-20240027). J. Mateu has been funded by research projects CIAICO/2022/191 from Generalitat Valenciana and PID2022-141555OB-I00 from the Spanish Ministry of Science and Innovation. The authors are grateful to Mehdi Moradi for providing useful sample code and advice.

\section*{Data Availability Statement}
The datasets and source code can be downloaded from \url{https://github.com/SujeongLeee/stdplinnet}.

\section*{Conflict of Interest Statement}
The authors declare that they have no conflicts of interest.

%%%%%%%%%%%%%%%%%%%%%%%%%%%%%%%%%%%%%
\clearpage
\bibliography{crime}

\end{document}

\maketitle
\section{The Bayesian hierarchical model structure}

\begin{align*}
    \textbf{Likelihood:} \quad & \prod_{i=1}^N \left[ \sum_{\substack{\forall j \text{ non-empty}}}
    q_j K_S(\mathbf{x}_i;\mathbf{c}_j^s, w_{s})K_T(t_i; c_{j}^{t}, w_{t}) \right]\\
    \textbf{Dirichlet process prior:} \quad & g_i \sim \text{Categorical}(q_1, \ldots, q_M)\\ 
    & q_j = U_1 \text{ for } j = 1 \text{ and } q_{j} = U_j \prod_{m=1}^{j-1} (1 - U_m) \quad \text{for } j = 2, \ldots, M \\
    & U_1, \ldots, U_m \overset{\text{iid}}{\sim} \text{Beta}(1, b_u)\\
    & b_u \sim \text{Gamma}(1, c)\\
    \textbf{Parameter prior:} \quad & w_s \sim N(\mu_s, 400)I_{[100,1000](w_s)}\\
    \quad & w_t \sim N(\mu_t, 400)I_{[0,\infty)}(w_t)\\
    \textbf{Cluster prior:} \quad & \mathbf{c} \sim \frac{1}{|\Delta|\mathcal{T}}
\end{align*}

\section{Full conditionals for MCMC}

%\textcolor{red}{why points 1,2,3 are repeated?}

\begin{enumerate}
  \item The conditional distribution of range parameters \((w_s, w_t)\):
    \begin{align*}
        \pi(w_s, w_t|\text{others})
        \propto
        & \prod_{i=1}^N f((\mathbf{x}_i, t_i)|w_s, w_t)\pi(w_s)\pi(w_t)\\
        \propto
        & \prod_{i=1}^N \left[ \sum_{\substack{\forall j \text{non-empty}}} q_j K_S(\mathbf{x}_i;\mathbf{c}_j^s, w_s)K_T(t_i; c_{j}^{t}, w_t) \right] \times \pi(w_s)\pi(w_t).\\
    \end{align*}
    
    \item The conditional distribution of membership variable \(g_i\):
    \[\pi(g_i|\text{others}) = \text{Categorical}(\tilde{q}_{i1},\dots,\tilde{q}_{iM})\]
    where
    \begin{align*}
        \tilde q_{ij} \propto q_j K_S(\mathbf{x}_i;\mathbf{c}_j^s, w_s)K_T(t_i; c_{j}^{t}, w_t)
    \end{align*}
    for $j = 1, \cdots, M.$\\
    
    \item The conditional distribution of space-time cluster centers \((\textbf{c}_j^s,c_j^t)\):
    \begin{align*}
        \pi((\textbf{c}_{j}^s,c_{j}^t)|\text{others}) \propto
        & \prod_{\substack{\forall i|g_i=j}}
        q_{j} K_S(\mathbf{x}_i;\mathbf{c}_j^s, w_s)K_T(t_i; c_{j}^{t}, w_t) \times \cfrac{1}{|\Delta|\mathcal{T}}.
    \end{align*}
    for $j = 1, \cdots, M.$\\
    
    \item The conditional distribution of \(U_{j}\):
    \[\pi(U_{j}|\text{others}) = \text{Beta}\left(1+\sum_{i=1}^{N} \mathbbm{1}\{g_i = j\}, b_u+\sum_{i=1}^{N} \mathbbm{1}\{g_i > j\}\right).\]\\
    
    \item The conditional distribution of \(b_{u}\):
    \[\pi(b_{u}|\text{others}) = \text{Gamma}\left(M-1+a, c-\sum_{j=1}^{M-1} \log(1-U_{j})\right),\]
    where a=1, c=1 and 1/4.
\end{enumerate}

% \clearpage
\section{Grid computation details} \label{grid}
\begin{figure}[h]
        \centering
            \makebox{\includegraphics[width=\linewidth]{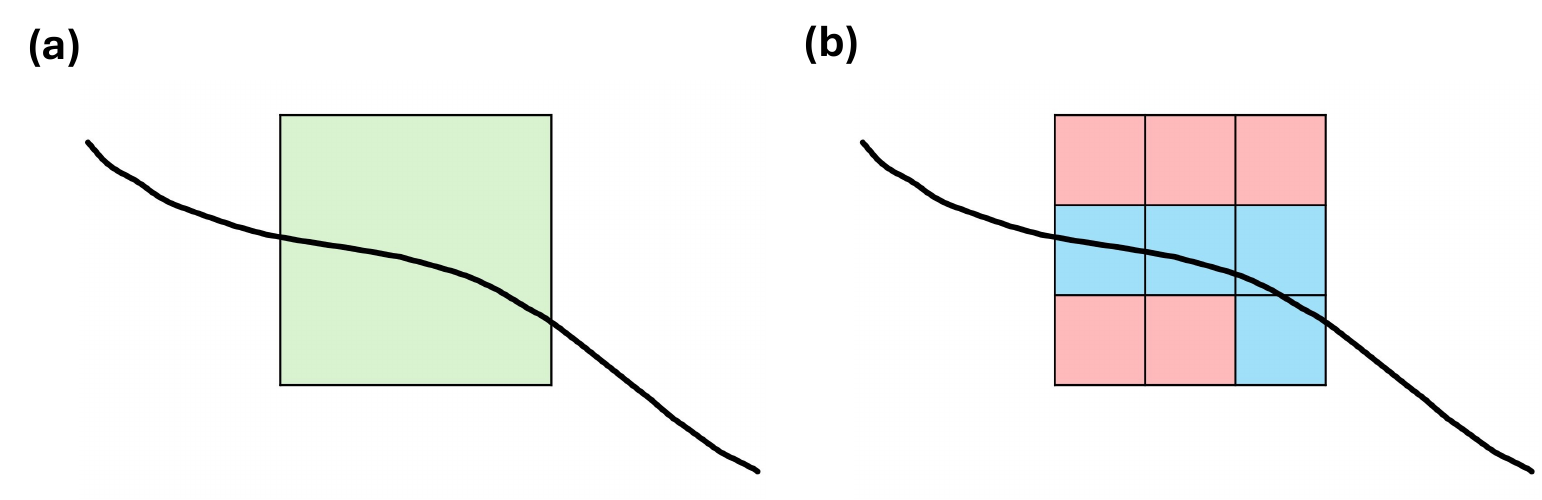}}
        \caption{
        (a) Grid and subgrid (b) covering the road network. Among the cubes belonging to the subgrid, only the part containing the road network (blue colors) was used in the aggregation step.}
        \label{fig:supp_grid}
\end{figure}

\clearpage
\section{Residual maps for model validation}

\begin{figure}[htbp]
    \centering
    \includegraphics[width=0.9\linewidth]{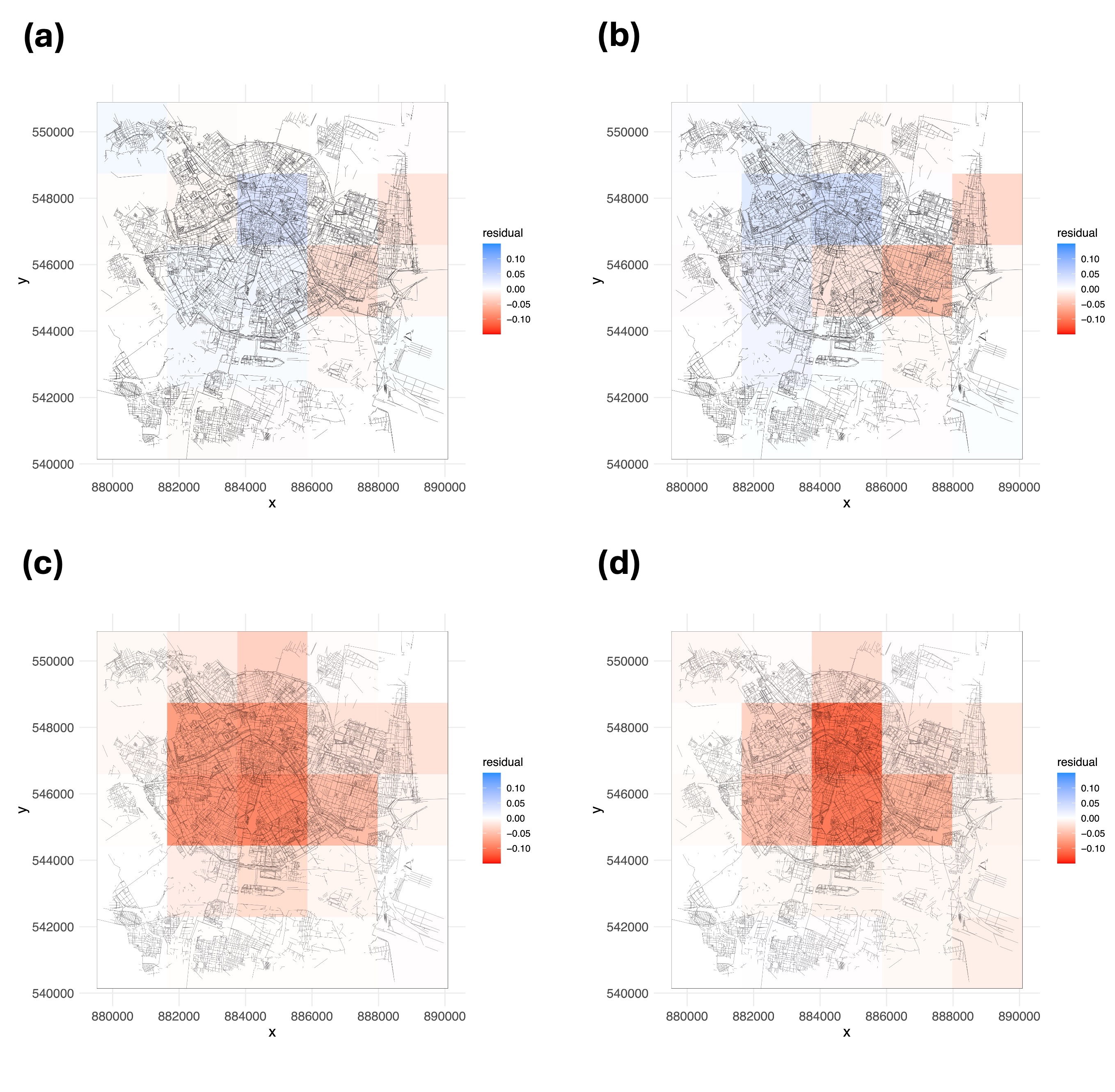}
    \caption{(a), (b) Residual maps of robbery and theft from the linear network DP model. (c), (d) Residual maps of robbery and theft from the standard DP model.}
	\label{fig:residualmap}
\end{figure}

We compute the residuals over a $G = 5 \times 5 \times 10$ grid, which is the same grid used for comparing observed and theoretical proportions. For visualization over the two-dimensional spatial domain, we aggregate the residuals by summing over the temporal grid points.
Figure~\ref{fig:residualmap} shows that the residuals from the linear network model are more distributed around zero, indicating a better fit compared to the standard DP model.

\section{Spatio-temporal conditional autoregressive model}

Let $y_{kt}$ and $x_{kt}$ denote the log-transformed robbery intensity and log-transformed theft intensity at location $k$ and time $t$, respectively. We use a $64\times64$ spatial grid, where each pixel is associated with an estimated intensity value at eight time points corresponding to quarterly intervals over two years. Therefore, the indices range are $k=1,\cdots, 4096$ and $t=1,\cdots,8$. We fit a spatio-temporal conditional autoregressive (CAR) model \citep{lee2021carbayesst} using the \texttt{ST.CARar} function from the \texttt{CARBayesST} package, defined as follows:
\begin{align*}
    \label{eq:stmodel}
    y_{kt} & \sim \mathcal{N}(\beta_0 +x_{kt}\beta_1 + \phi_{kt}, \nu^2)\\
    \boldsymbol \phi_t|\boldsymbol \phi_{t-1} &\sim \mathrm{N}\left( \rho_T \boldsymbol \phi_{t-1}, \, \tau^2 \mathbf{Q}(\mathbf{W}, \rho_S)^{-1} \right)\\
    \boldsymbol \phi_1 &\sim \mathrm{N}\left( \mathbf{0}, \, \tau^2 \mathbf{Q}(\mathbf{W}, \rho_S)^{-1} \right)\\
    \tau^2 &\sim \mathrm{Inverse\text{-}Gamma}(a, b)\\
    \rho_S, \rho_T &\sim \mathrm{Uniform}(0, 1),
\end{align*}
where $\mathbf W = (w_{kj})$ is a $K \times K$ adjacency matrix, and $\phi_{kt}$ is a spatio-temporal random effect at location $k$ and time $t$. The vector $\boldsymbol \phi_t=(\phi_{1t},\cdots,\phi_{4096t})$ represents the spatial random effects at time $t$. The precision matrix is defined as $\mathbf Q(\mathbf W, \rho_S) = \rho_S[\text{diag}(\mathbf{W1})-\mathbf{W}]+(1-\rho_S)\mathbf{I}$, where $\mathbf 1$ is a $K \times 1$ vector of ones, and $\mathbf I$ is the $K \times K$ identity matrix. $\nu^2$ denotes the error variance, $\tau^2$ is the variance of the spatio-temporal random effects, and $\rho_S$ and $\rho_T$ are the spatial and temporal autocorrelation coefficients. We run the MCMC algorithm for 400,000 iterations to fit the model, and the results are summarized in Table~\ref{tab:fig2streg}. We observe that the regression coefficient $\beta_1$ is significantly positive, indicating that theft has a positive impact on robbery. Furthermore, both $\rho_S$ and $\rho_T$ are statistically significant, suggesting the presence of spatial and temporal autocorrelation in the relationship between the two crime types.

\begin{table}[htbp]
\centering
\begin{tabular}{lccc}
\hline
Parameter & Mean & 2.5\% & 97.5\% \\
\hline
$\beta_0$ & -9.1445 & -9.1805 & -9.0994 \\
$\beta_1$ &  0.1277 &  0.1246 &  0.1316 \\
$\nu^2$ &  0.0001 &  0.0001 &  0.0001 \\
$\tau^2$ &  0.0996 &  0.0978 &  0.1015 \\
$\rho_S$ &  0.9999 &  0.9999 &  1.0000 \\
$\rho_T$ &  0.6906 &  0.6801 &  0.7012 \\
\hline
\end{tabular}
\caption{Posterior means and 95\% HPD intervals of the parameters.}
\label{tab:fig2streg}
\end{table}

\bibliography{crime.bib}